\documentclass[]{aastex631}

\shorttitle{\texttt{TTTT}}
\shortauthors{Gorodetsky, Mullen, et. al.}

\usepackage{xcolor}
\usepackage{todonotes}
\usepackage{amsmath}
\usepackage[ruled,vlined,linesnumbered]{algorithm2e}
\usepackage{enumitem}

\makeatletter
\def\blfootnote{\gdef\@thefnmark{$\dagger$}\@footnotetext}
\makeatother

\newcommand{\reals}{\mathbb{R}}
\newcommand{\naturals}{\mathbb{N}}
\newcommand{\cx}[1]{#1_x}
\newcommand{\ct}[1]{#1_{\theta}}
\newcommand{\cp}[1]{#1_{\phi}}

\begin{document}

\title{Thermal Radiation Transport with Tensor Trains}

\author{Alex A. Gorodetsky$^\dagger$}
\affiliation{Department of Aerospace Engineering, University of Michigan, Ann Arbor, MI}
\affiliation{Michigan SPARC, Los Alamos National Laboratory, Ann Arbor, MI}

\author[0000-0003-2131-4634]{Patrick D. Mullen$^\dagger$}
\affiliation{Michigan SPARC, Los Alamos National Laboratory, Ann Arbor, MI}
\affiliation{Computational Physics and Methods, Los Alamos National Laboratory, Los Alamos, NM}

\author{Aditya Deshpande}
\affiliation{Department of Aerospace Engineering, University of Michigan, Ann Arbor, MI}
\affiliation{Michigan SPARC, Los Alamos National Laboratory, Ann Arbor, MI}

\author[0000-0003-4353-8751]{Joshua C. Dolence}
\affiliation{Michigan SPARC, Los Alamos National Laboratory, Ann Arbor, MI}
\affiliation{Computational Physics and Methods, Los Alamos National Laboratory, Los Alamos, NM}
\affiliation{Department of Aerospace Engineering, University of Michigan, Ann Arbor, MI}

\author[0000-0002-7530-6173]{Chad D. Meyer}
\affiliation{Michigan SPARC, Los Alamos National Laboratory, Ann Arbor, MI}
\affiliation{Continuum Models and Numerical Methods, Los Alamos National Laboratory, Los Alamos, NM}

\author[0000-0001-6432-7860]{Jonah M. Miller}
\affiliation{Michigan SPARC, Los Alamos National Laboratory, Ann Arbor, MI}
\affiliation{Computational Physics and Methods, Los Alamos National Laboratory, Los Alamos, NM}

\author[0000-0001-7364-7946]{Luke F. Roberts}
\affiliation{Michigan SPARC, Los Alamos National Laboratory, Ann Arbor, MI}
\affiliation{Computational Physics and Methods, Los Alamos National Laboratory, Los Alamos, NM}








\begin{abstract}
\blfootnote{Both co-lead authors provided equal contributions.}
We present a novel tensor network algorithm to solve the time-dependent, gray thermal radiation transport equation.  The method invokes a tensor train (TT) decomposition for the specific intensity. The efficiency of this approach is dictated by the rank of the decomposition. When the solution is ``low-rank,'' the memory footprint of the specific intensity solution vector may be significantly compressed.  The algorithm, following a step-then-truncate approach of a traditional discrete ordinates method, operates directly on the compressed state vector thereby enabling large speedups for low-rank solutions. To achieve these speedups we rely on a recently developed rounding approach based on the Gram-SVD.   We detail how familiar $S_N$ algorithms for (gray) thermal transport can be mapped to this TT framework and present several numerical examples testing both the optically thick and thin regimes. The TT framework finds low rank structure and supplies up to $\simeq$60$\times$ speedups and $\simeq$1000$\times$ compressions for problems demanding large angle counts, thereby enabling previously intractable $S_N$ calculations and supplying a promising avenue to mitigate ray effects.
\end{abstract}



\section{Introduction} \label{sec:intro}

Thermal radiation transport (i.e., photon radiative transfer coupled to a background fluid) often determines the thermodynamics and evolution of (1) astrophysical environments/phenomena, such as black hole accretion disks \cite[see, e.g.,][and references therein]{mckinney14,jiang14_super,sadowski14,jiang2019,huang23_bh,liska23,asahina24}, disk winds \citep[e.g.,][]{proga2000,proga2004,dyda23}, neutron star accretion columns \citep{zhang21,zhang22,zhang23,sheng23}, tidal disruption events \citep[e.g.,][]{huang23, huang24}, proto-Jupiter convective envelopes and protoplanetary disks  \citep[e.g.,][]{flock13,flock16,zhu21,zhang24,krapp24}, and stellar interiors \citep[see][for a review]{jiang23}, to name a few, as well as (2) high energy density physics (HEDP) platforms, such as inertial confinement fusion experiments \citep[see, e.g.,][]{haines22,haines23,kim23,ko24}.

The time-dependent, photon radiative transfer equation evolves the specific intensity $I_\nu \equiv I_\nu(t,\vec{\mathbf{x}},\vec{\mathbf{\Omega}})$. This function is \textit{seven} dimensional: a function of time $t$, frequency $\nu$, three spatial coordinates $\vec{\mathbf{x}}$, and two angular coordinates $\vec{\mathbf{\Omega}}$ defining the direction of propagation.  A common approach to numerically solving the transport equation are the discrete ordinates ($S_N$) methods.  In $S_N$-like schemes, the specific intensity is directly discretized via $N_x \times N_y \times N_z$ spatial cells (in 3D), each containing $N_\mathrm{angles}$ angular ordinates, and $N_\mathrm{freq}$ frequency bins (e.g., ``multigroup").  Such methods therefore suffer from the ``curse of dimensionality," as the solution is comprised of $N_x N_y N_z N_\mathrm{angles} N_\mathrm{freq}$ unknowns.  Often, a large number of spatial cells ($N_x N_y N_z$) are required to resolve complicated spatial features. Likewise, many frequency bins ($N_\mathrm{freq}$) are often required due to the complex frequency-dependent structure of material opacities.  To mitigate ``ray-effects''---wherein specific intensity is preferentially transported along directions associated with the angular discretization---large $N_\textrm{angles}$ is also often required. As a result, standard approaches to $S_N$ can become prohibitively expensive in many contexts.

A number of alternative methods for solving the transport equation exist. $P_N$ methods use a spherical harmonic basis to represent the specific intensity over the angular domain; such methods are subject to ``wave effects" and struggle with maintaining positivity \citep{brunner02}.  Monte Carlo methods randomly sample the radiation field via $N_\mathrm{p}$ particles (which are individually tracked, undergoing transport steps as well as scattering and/or absorption events) and are subject to Poisson noise $\propto 1/\sqrt{N_\mathrm{p}}$ \citep[][]{imc71}.  Moment-based methods evolve a truncated number of moments of the radiation field (e.g., the radiation energy density, flux, etc.), hence requiring a closure model.  M1 methods \citep[][]{m1}, which close the evolved first two moments via a prescribed radiation pressure tensor, struggle with crossing beams tests \cite[i.e., by unphysically merging beams, see tests in][]{foucart15}.  Variable Eddington tensor (VET) methods supply closures via short characteristics \citep{davis2012} or particle swarms \citep{mocmc,IzquierdoGuidedMoments} and are therefore again subject to angular errors or noise, respectively.  Diffusion models evolve the energy density directly via a simple parabolic equation, but are not applicable in the optically thin regime and often require modifying the solved equations to avoid unphysical behavior \citep[e.g., flux-limited diffusion,][]{levermore81}.  A number of methods have been proposed to directly mitigate ray effects: e.g., reference frame rotations \citep{tencer16,camminady19} or hybrid $S_N/P_N$ methods and finite element approaches \citep{lathrop71,geo_ray23}.  These methods either (a) require augmenting existing algorithms with additional quadrature rotations and solution interpolations or (b) invoke limiters/filters to preserve positivity.

We propose to combat the curse of dimensionality in thermal transport by expressing ``traditional" $S_N$ algorithms in tensor train (hereafter, TT) format \citep{oseledets2011tensor}. The TT format reduces the storage complexity of a $d$ dimensional array of size $N$ from $N^d$ to one that is $\mathcal{O}(dNr^2).$ The exponential complexity is reduced to a complexity dictated by the rank $r$ of the representation. When solutions are low rank, substantial gains in both compression and solver speed can be obtained. Low rank TT decompositions have been previously considered for Boltzmann-like equations \citep[see][for a review]{einkemmer24}, however, in this work we specialize to gray thermal transport with tensor trains (hereafter, \texttt{TTTT}) and applications therein. We employ a step and truncate approach that has been previously applied for many other equations including the Hamilton-Jacobi-Bellman equation \citep{gorodetsky2018high}, Fokker-Planck equation \citep{dolgov2012fast}, Vlasov-Poisson Equation \citep{kormann2015semi}, and Schrodinger equation \citep{soley2021functional}. The main idea of these approaches is that the operators and source terms for these equations have a natural low-rank structure that can be exploited to analytically apply them to a solution. These applications can then be assembled in the context of ``tensorfying'' existing numerical methods by performing all required arithmetic in tensor format, and re-truncating the solution to a guaranteed tolerance after this application. We contrast these types of approaches with those that propagate the tensor factors directly through equations separately via specifically tailored numerical methods \citep{koch2007dynamical,lubich2013dynamical, ceruti2022rank,dektor2021dynamic} to combine them. We refer to the aforementioned review paper and citations for a discussion of the benefits and challenges of each approach.

In this paper, we focus on developing and assessing \texttt{TTTT} in the simplest interesting context, namely gray transport in static media in two spatial dimensions.  The rest of this paper is structured as follows. In \S \ref{sec:governing_eqs}, we introduce the photon transport equation and the material temperature equation it couples to.  In  \S \ref{sec:discretization}, we introduce the discretization adopted in this work---a time-explicit variant of the \cite{jiang21} algorithm.  In \S \ref{sec:tensors}, we introduce our TT decomposition and describe operations (e.g., addition, multiplication, integration) in this tensor format.  In \S \ref{sec:tests}, we present common tests in the thermal transport literature at high angular fidelity.  For problems requiring large $N_\mathrm{angles}$, we demonstrate that \texttt{TTTT} finds significant compressions (up to $\simeq$1000$\times$) and/or large speedups (up to $\simeq$60$\times$). We provide discussions and conclusions in \S \ref{sec:discussion} and \S \ref{sec:conclusions}, respectively, highlighting tensors as a viable, exciting pathway towards mitigating ray effects and enabling previously impossible calculations in thermal radiation transport. 

\section{Governing Equations} \label{sec:governing_eqs}
The gray thermal transport equation
\begin{equation}
\partial_t I + c \mathbf{n} \cdot \nabla I = c \left(j - \alpha I \right)
\label{eq:gray}
\end{equation}
evolves the frequency-integrated specific intensity $I=\int_0^\infty I_\nu d \nu $, where $c$ is the speed of light, $\mathbf{n}$ indicates a unit normal coordinate direction, and $j$ and $\alpha$ are the frequency-integrated emissivity and absorptivity, respectively. The radiation energy density $E$ is related to the zeroth moment of the specific intensity (i.e., $J \equiv$ the mean intensity) via 
\begin{equation} 
    E = \frac{4 \pi}{c} J = \frac{1}{c} \int I d \Omega,
    \label{eq:energy_density}
\end{equation}
where $d\Omega$ is differential solid angle.  In this work, we (1) assume local thermodynamic equilibrium (LTE) and hence the gray form of Kirchoff's law, (2) invoke fluid-frame isotropic, elastic scattering, and (3) mandate that the background medium with material density $\rho$ and temperature $T$ is static (i.e., $\mathrm{material \; velocity} = 0$), thereby obviating the requirement for frame transformations \citep[see, e.g.,][]{jiang21, white23} or, more generally, any material-motion corrections \citep[see, e.g.,][]{lowrie23}. Under these assumptions, we can recast Equation (\ref{eq:gray}) as
\begin{equation}
    \partial_t I + c \mathbf{n} \cdot \nabla I = c \left [\rho \kappa_s \left(J - I \right) + \rho \kappa_a \left(\frac{c}{4 \pi} a_r T^4 - I \right) \right],
\label{eq:transport_tt}
\end{equation}
where $a_r$ is the radiation constant, and the first and second terms of the coupling expression account for scattering (subject to specific scattering opacity $\kappa_s$) and thermal emission/absorption (subject to specific absorption opacity $\kappa_a$), respectively.

Total energy conservation mandates that the change in the radiation energy density $E$ due to the radiation source term matches the change in the material internal energy density $u$.  Assuming an ideal equation of state with $u = \rho c_v T$, where $c_v$ is a constant specific heat capacity (at constant volume), integration of the right hand side of Equation (\ref{eq:transport_tt}) over solid angle gives a corresponding material temperature equation
\begin{equation}
        \partial_t T = - \frac{c \kappa_a}{c_v} (a_r T^4 - E).
\label{eq:temp_eq}
\end{equation}

\section{Discretization} \label{sec:discretization}
We discretize the gray specific intensity as $I^{m}_{i \ell p}$, where $m$ is a time index, $i$ is a spatial index mapping logical space to cell-centered locations on a uniform Cartesian mesh, and the $\ell, p$ pair index polar and azimuthal angle centers $\theta$ and $\phi$ in our latitude-longitude discretization of angular space, respectively. Our angular discretization (1) uniformly samples $\cos \theta$ and $\phi$, (2) ensures that $\sum_{\ell,p} \Delta \Omega_{\ell p} = 4 \pi$, where $\Delta \Omega_{\ell p}$ is the subtended solid angle of angle $\ell,p$, and (3) is a 2D tensor product grid (see later \S \ref{sec:tensors}).  Angle unit normals are defined by $\mathbf{n} = \left[n_x, n_y, n_z \right] = \left[ \sin \theta_\ell \cos \phi_p, \sin \theta_\ell \sin \phi_p, \cos \theta_\ell \right]$.  The discretized radiation energy density is 
\begin{equation}
    E^{m}_{i} = \frac{1}{c} \sum_{\ell ,p} I^{m}_{i \ell p} \Delta \Omega_{\ell p}.
\end{equation}
We similarly discretize the density $\rho^{m}_{i}$, temperature $T^{m}_{i}$, specific heat $c^{m}_{v,i}$, and specific opacities $\kappa^{m}_{a,i}$ and $\kappa^{m}_{s,i}$.  In this work, we restrict ourselves to one and two spatial dimensions, though extensions to 3D are straightforward.

\subsection{Transport Operator} \label{sec:transport}
We invoke a first-order time-explicit (RK1), finite volume construction for the transport operator that can conservatively advance the cell-volume-averaged specific intensity $I_{i \ell p}$ via the divergence of area-averaged, interface fluxes $\mathbf{F} = c \mathbf{n} I$.  In 1D, where the index $i$ corresponds to direction $x$, we have
\begin{equation}
    \frac{I^{m+1}_{i \ell p} - I^{m}_{i \ell p}}{\Delta t} + \frac{F^{m}_{i + 1/2, \ell p} - F^{m}_{i - 1/2, \ell p}}{\Delta x} = 0,
    \label{eq:lhs}
\end{equation}
where half integers are used to denote evaluations at cell faces and $\Delta t \leq \mathrm{CFL} \cdot \Delta x/c$ is the hyperbolic time step.  Various numerical constructions of $F_{i+1/2}$ exist for such hyperbolic systems. To demonstrate the utility of tensors (see later \S \ref{sec:tensors}), we examine (1) Rusanov, (2) upwind, and (3) HLL fluxes in this work. To keep the transport operator linear, we only consider piecewise constant reconstructions of the specific intensity, such that the left and right interface states (at $i+1/2$) are $U_L = I_{i \ell p}$ and $U_R = I_{i+1, \ell p}$, thereby giving fluxes $F_L = c n_x I_{i \ell p}$ and $F_R = c n_x I_{i+1, \ell p}$.

\paragraph{Rusanov}
\cite{toro2009} presents the Rusanov flux as
\begin{equation}
    F_{i+1/2} = \frac{1}{2} \left(F_L + F_R \right) - \frac{1}{2} S^+ (U_R - U_L),
\label{eq:rusanov_flux_toro}
\end{equation}
where $S^+$ is an extremal wavespeed.  In the optically thin regime, we select $S^{+} \sim c$ for stability, which results in significant numerical diffusion.  For optically thick problems, $S^{+}$ could be informed by the local optical depth $\tau_c$ (see later \S \ref{sec:hll}).  As $\tau \rightarrow \infty$, one could select $S^{+} \sim 0$ which returns a central difference scheme.

\paragraph{Upwinding}
Another strategy to advance the transport operator is via upwinding, wherein the interface flux is constructed via
\begin{equation}
    F_{i+1/2} = \left\{
\begin{array}{cc}
  F_L & \text{ if } n_x \geq 0\\
  F_R & \phantom{ \ if } n_x < 0.
  \end{array}   
\right. 
\label{eq:upwind}
\end{equation}
Upwinding works well in optically thin media, but demands extremely fine linear resolutions when $\tau \gg 1$ to recover the diffusive regime.

\paragraph{HLL} \label{sec:hll}
\cite{toro2009} gives the HLL flux as
\begin{equation}
    F_{i+1/2} = \frac{S_R F_L - S_L F_R + S_L S_R (U_R - U_L)}{S_R - S_L}.
    \label{eq:hll}
\end{equation}
\cite{jiang21} defines expressions for wavespeeds $S_{L,R}$ that attempt to capture asymptotic behavior:
\begin{equation}
    S_{R,i+1/2} = \left\{
\begin{array}{cc}
  c n_x \sqrt{\left[1 - \exp \left(- \tau_{c, i+1/2}^2 \right) \right] / \tau_{c, i+1/2}^2} & \text{ if } n_x \geq 0 \; \mathrm{and} \; \tau_{c, i+1/2} \geq \tau_\mathrm{threshold} \\
  c n_x \sqrt{1 - \tau_{c, i+1/2}^2 / 2} & \phantom{ if } n_x \geq 0 \; \mathrm{and} \; \tau_{c, i+1/2} < \tau_\mathrm{threshold} \\
  -c n_x \sqrt{\left[1 - \exp \left(- \tau_{c, i+1/2}^4 \right) \right] / \tau_{c, i+1/2}^2} & \phantom{ if } n_x < 0 \; \mathrm{and} \; \tau_{c, i+1/2} \geq \tau_\mathrm{threshold} \\
  - c n_x \tau_{c, i+1/2} & \phantom{ if } n_x < 0 \; \mathrm{and} \; \tau_{c, i+1/2} < \tau_\mathrm{threshold}
  \end{array} 
\right.
\end{equation}

\begin{equation}
    S_{L,i+1/2} = \left\{
\begin{array}{cc}
  -c n_x \sqrt{\left[1 - \exp \left(- \tau_{c, i+1/2}^4 \right) \right] / \tau_{c, i+1/2}^2} & \text{ if } n_x \geq 0 \; \mathrm{and} \; \tau_{c, i+1/2} \geq \tau_\mathrm{threshold} \\
  - c n_x \tau_{c, i+1/2} & \phantom{ if } n_x \geq 0 \; \mathrm{and} \; \tau_{c, i+1/2} < \tau_\mathrm{threshold} \\
  c n_x \sqrt{\left[1 - \exp \left(- \tau_{c, i+1/2}^2 \right) \right] / \tau_{c, i+1/2}^2} & \phantom{ if } n_x < 0 \; \mathrm{and} \; \tau_{c, i+1/2} \geq \tau_\mathrm{threshold} \\
  c n_x \sqrt{1 - \tau_{c, i+1/2}^2 / 2} & \phantom{ if } n_x < 0 \; \mathrm{and} \; \tau_{c, i+1/2} < \tau_\mathrm{threshold}.
  \end{array} 
\right.
\end{equation}
In the limit of $\tau \rightarrow 0$, the upwind flux is recovered.  A series expansion of the asymptotic preserving flux about $[\tau_c]^{-1} = 0$ recovers the Rusanov flux with $S^{+} = c \lvert n_x \rvert / \tau_c$.  In this work, we define the local optical depth via a harmonic mean
\begin{equation}
    \tau_{c, i+1/2} = \frac{2 \beta \Delta x}{(\rho_i \kappa_{a,i} + \rho_i \kappa_{s,i})^{-1} + (\rho_i \kappa_{a,i+1} + \rho_i \kappa_{s,i+1})^{-1}},
\label{eq:tauc}
\end{equation}
where $\beta$ and $\tau_\mathrm{threshold}$ are tuneable constants.  

\subsection{Absorption, Emission, and Scattering} \label{sec:rhs}
Despite evolving the transport operator explicitly, we must adopt an implicit discretization for absorption, emission, and scattering, since their associated timescales in the optically thick regime ($t_a = [c \rho \kappa_a]^{-1}$ and $t_s = [c \rho \kappa_s]^{-1}$) can be significantly shorter than the light-crossing time of the numerical cell.  Therefore, in a backward Euler, operator-split update, we seek to solve the system
\begin{eqnarray}
    \frac{I^{m+1}_{i \ell p} - I^{m}_{i \ell p}}{\Delta t} & = & c \left [ \rho^{m}_{i} 
 \kappa^{m}_{s,i} \left(J^{m+1}_{i} - I^{m+1}_{i \ell p} \right) + \rho^{m}_{i} 
 \kappa^{m}_{a,i} \left(\frac{c}{4 \pi} a_r (T^\dagger_{i})^4 - I^{m+1}_{i \ell p} \right) \right], \label{eq:couple_a} \\
    \frac{T^\dagger_{i} - T^m_{i}}{\Delta t} & = & - \frac{c \kappa^m_{a,i}} {c_{v,i}} \left( a_r (T^\dagger_{i})^4 - \frac{4 \pi}{c} J^{m+1}_{i} \right) \label{eq:couple_b} 
\end{eqnarray}
where we have assumed that $\kappa_a$ and $\kappa_s$ are constant over the coupling stage.  Solving Equation (\ref{eq:couple_a}) for $I^{m+1}_{i \ell p}$ and integrating over solid angle gives an expression for $J^{m+1}_{i}$ that can be substituted into Equation (\ref{eq:couple_b}).  Equation (\ref{eq:couple_b}) can then be solved for $T^\dagger_{i}$ via a simple fourth order polynomial root find \citep[c.f.,][]{jiang21}.  Once $T^\dagger_{i}$ is identified, $I^{m+1}_{i \ell p}$ is straightforwardly calculable.  Rather than directly assigning $T^{m+1}_{i} = T^\dagger_{i}$, we instead set $T^{m+1}_{i}$ via total energy conservation:
\begin{equation}
    \rho_{i} c_{v,i} (T^{m+1}_{i} - T^{m}_{i}) = \frac{4 \pi}{c} \left( J^{m}_{i} - \frac{1}{4 \pi} \sum_{\ell,p} I^{m+1}_{i \ell p} \Delta \Omega_{\ell p} \right).
\end{equation}

\section{Tensors} \label{sec:tensors}

\subsection{Tensor solution ansatz and solution architecture}\label{sec:tensor_ops}
In 3D, the gray specific intensity is described by $N_xN_yN_zN_\theta N_\phi$ unknowns. Our approach for representing this solution is to group all of the spatial variables together and to separate out each of the angular variables individually. To this end, we will consider the overall tensor as having  $N_x N_{\theta} N_{\phi}$ unknowns, were $N_x$ should be understood to represent the total number of spatial unknowns through an abuse of notation. We choose this approach, rather than separating the spatial dimensions, so that we can handle arbitrary geometries and potentially non-tensor product meshes in future work.

The particular separation that we follow in this paper is  called the tensor train \citep{oseledets2011tensor} format. More specifically, we represent the specific intensity tensor as a product of three {\it TT-cores}:
\begin{equation}
 \hat{I}^m_{i \ell p} = X^m_{i} \Theta^m_{\ell} \Phi^m_{p} 
\label{eq:tensor_decomp}
\end{equation}
where $X^m \in \reals^{N_x \times r_1^m}$ represents the spatial variables, $\Theta \in \reals^{r_1^m \times N_\theta \times r_2^m}$, and $\Phi^m \in \reals^{r_2^m \times N_\phi}$, for all $i,\ell,p$. The sizes $r_1^m, r_2^m \in \naturals$ are termed the {\it ranks} of this representation. Colloquially, the rank of the representation encodes the amount of information required to represent it. The storage requirement of the 1D specific intensity tensor in TT format is $r_1^m N_x + r_1^m r_2^m N_\theta +r_2^m N_\phi$. While the description that follows is for 1D spatial meshes, we apply virtually the same operations to the multi-dimensional case by grouping all spatial dimensions together. For example, for a 2D problem, the first core would have size $X^m \in  \reals^{N_xN_y \times r_1^m}.$ When indexing cores, $X_i^m$ refers to the $i$-th row of $X^m$, $\Theta_{l}^m$ refers to an $r_1^m \times r_2^m$ matrix, and $\Phi_p^m$ refers to the $p$-th column of $\Phi$.

In this work, we present a solver that is constructed from operations on the solution represented in this multi-linear ansatz. Furthermore, the solver can leverage this multi-linearity to accelerate certain operations.

\paragraph{Addition}
Consider two three-dimensional tensors $T_{i \ell p}$ and $S_{i \ell p},$ with corresponding cores $(T_1,T_2,T_3)$ and $(S_1,S_2,S_3)$, respectively. Addition of tensors in TT format yields a tensor $V_{i \ell p}$ with the following format
\begin{align}
    V_{i \ell p} &= T_{i \ell p} + S_{i \ell p} = 
    \underbrace{\begin{bmatrix}
    T_{1,i} & S_{1,i}
    \end{bmatrix}}_{V_{1,i}}
    \underbrace{
    \begin{bmatrix}
        T_{2,\ell} & 0 \\
        0 & S_{2,\ell}
    \end{bmatrix}
    }_{V_{2,\ell}}
    \underbrace{
    \begin{bmatrix}
    T_{3,p} \\
    S_{3,p}
    \end{bmatrix}
    }_{V_{3,p}}
    = V_{1,i}V_{2,\ell}, V_{3,p},
\end{align}
with the ranks of $V_{i \ell p}$ being the {\it sum} of the corresponding ranks of $T$ and $S$. Moreover, we see that this additive structure yields a block-diagonal structure for the central core.  

\paragraph{Integration}
Integration is an equally efficient operation in tensor format. Consider a tensor-product numerical integration rule (e.g., for the radiation energy density) where we seek to integrate over a subset of the dimensions according to 
\begin{equation}
E_{i} = \frac{1}{c}\sum_{\ell p} I_{i \ell p} w_{\ell} v_p,
\end{equation}
where $\{w_{\ell}\}$ denote the quadrature weights over $\theta$ and $\{v_p\}$ denote the weights over $\phi$. If $I$ is in tensor train format, this integral readily becomes 
\begin{equation}
    E_{i} =  \frac{1}{c} X_i \left(\sum_{\ell} \Theta_\ell w_\ell  \right)\left(\sum_p \Phi_p v_p \right).
\label{eq:tt_energy_density}
\end{equation}

\paragraph{Multiplication}
Element-wise multiplication of two tensors in TT format is also readily computed. Consider the same $T_{ijk}$ and $S_{ijk}$ as for addition. Then the tensor $V_{ijk} = T_{ijk} S_{ijk}$ is given by 
\begin{align}
V_{ijk} &= T_{ijk}S_{ijk} = \left[ \underbrace{\cx{T}[i] \otimes \cx{S}[i]}_{\cx{V}[i]} \right]
\left[ \underbrace{\ct{T}[j] \otimes \ct{S}[j]}_{\ct{V}[j]} \right]
\left[ \underbrace{\cp{T}[k] \otimes \cp{S}[k]}_{\cp{V}[k]} \right] = \cx{V}[i]\ct{V}[j] \cp{V}[k],
\end{align}
where $\otimes$ refers to the outer product. Here, the ranks of $V_{ijk}$ are the {\it product} of the ranks of $T$ and $S$. Thus this operation is considerably more expensive than addition. 

\subsection{Rounding}
    Basic TT-operations for addition and multiplication inflate the ranks of our solution when performed exactly.  As a result, the ranks of the solution vector can grow uncontrollably with each timestep without any mitigation strategy. 
    Rounding algorithms are used to mitigate this rank growth by re-approximating a tensor with one of lower rank to a fixed error tolerance.  
    
    Briefly, different rounding approaches all fundamentally rely on different low-rank approximations to a matrix-matrix product (essentially, rounding a matrix-matrix product). 
    In this setting, given two matrices $A B^T$ where, $A \in R^{m \times r}$ and $B \in R^{m \times r}$, the goal is to find a  low-rank approximation $A' B'^T$ where $A' \in R^{m \times r'}$ and $B' \in R^{n \times r'}$ such that $r' < r$ where the quality of the approximation, quantified by the norm (usually Frobenius) of the error, is maintained to be $|| A B^T - A'B'^T ||\leq \epsilon$ for some user-specified tolerance $\epsilon.$  The TT can be viewed as a sequence of matrix-matrix multiplications and so the TT rounding problem can be formulated as a sweeping algorithm that considers each dimension separately. Different approaches for this low-rank approximation yield different TT rounding algorithms. We provide interested readers details and further references in Appendix \ref{sec:appendix_rounding}. The approach in this work is to round after each transport update and after each operator split update via the GRAM-SVD algorithm \citep{al2022parallel}. We have found this algorithm to speed up the computation up to five times compared with the standard TT-SVD approach \citep{oseledets2011tensor}.

\subsection{Transport Operator}
Spatial components of the angle normals
\begin{align}
n_x &= \sin \theta \cos \phi \\
n_y &= \sin \theta \sin \phi
\end{align}
are both rank-1 tensors. The fluxes $F_L$ and $F_R$ are therefore obtained by multiplication of rank-1 tensors with the TT representation of the specific intensity tensor; furthermore, these representations do not change ranks. Using these terms, we find that the explicit update of Equation (\ref{eq:lhs}) requires the addition of three tensors $I^m$, $-\frac{\Delta t}{\Delta x} F^m_{i+1/2}$ and $+\frac{\Delta t}{\Delta x}F^m_{i-1/2}$.

\paragraph{Rusanov} 
The Rusanov flux~\eqref{eq:rusanov_flux_toro} is just the sum of 4 tensors, thus increasing the rank to $4r$.
\paragraph{Upwind}
The upwind flux has a switch condition that is entirely a function of $\phi$. As a result, this can be formulated via a rank 1 masking tensors $M^+_x$ and $M^-_x$. For example, to mask for cases where $\cos(\phi) > 0$, the masking tensor is the rank one tensor $M^+_x = \mathbf{1}_{\cos \phi>0}$ where $\mathbf{1}$ is the indicator function. Under these conditions we have that $F_{i+1/2}$ is a sum of products of the specific intensity with the rank-1 masking tensors
\begin{equation}
F_{i+1/2} = F_L M^+_x + F_R M^-_x.
\end{equation}
Note that for multi-dimensional problems we need to mask for each spatial direction separately (e.g., the flux term ends up being a sum of four tensors in 2D).

\paragraph{HLL}

For the HLL flux, we first view $S_L$ and $S_R$ as rank 2 tensors formed by a similar masking operation as required by the upwind flux to handle the $n_x\geq0$ and $n_x <0$ cases (with similar considerations for higher-dimensional settings).  Similarly, we form $\frac{1}{S_R - S_L}$ by itself as a tensor using the same sort of masking procedure.  In practice, we could avoid an extra multiplication by directly forming $\frac{S_R}{S_R-S_L}$, $\frac{S_L}{S_R-S_L}$ and $\frac{S_RS_L}{S_R-S_L}$ first. Then, Equation (\ref{eq:hll}) is the sum of four products of tensors. These sums and products are all performed in TT format. 

\subsection{Operator splitting}
We solve the temperature equation in non-tensor form. The specific intensity enters this equation only through its moments. Such integrals are readily computed as shown in \S\ref{sec:tensor_ops}.  Once the temperature equation is solved, the source terms can be added to the specific intensity tensor by direct summation (see Equation \ref{eq:transport_tt}), and multiplications in their construction are all rank 1 because this work assumes angle-independent opacities.

\subsection{Boundary conditions}
We use a standard ghost cell approach for boundary conditions, but this approach needs to be tailored to the fact that a tensor is indirectly represented by its TT cores. Ghost cells are added only along the spatial dimensions, so this corresponds to adding rows to the top and bottom of $X^m_i$, prior to performing tensor operations.  Let the function $\mathtt{ghostpad}(X^m)$ indicate the addition of rows filled with zeros.

For Dirichlet boundary conditions we require directly setting elements of e.g., $I_{-1\ell p}$ to a given value $L$. To avoid modifications to the internal nodes, this must be executed via the addition of a rank one term. Specifically, we define the vector $Le_{-1}$ to denote a vector with $L$ in the $-1$ location (first ghost cell), and zero elsewhere. Then we define the rank one tensor required for the Dirichlet update as
\begin{equation}
D_{ip\ell} = [Le_{-1}]_i .
\end{equation}
Thus the Dirichlet update performs $\mathtt{ghostpad}(X^m) + D$ prior to each iteration. 

Periodic and outflow boundary conditions do not require any rank one updates. Periodic boundary conditions are obtained by first adding a row to the bottom of the core $X^m$ and then setting it equal to the first row of  $X^m$.  Outflow boundary conditions are obtained by running the $\mathtt{ghostpad}(X^m)$ operation followed by setting the -1 row of $X^m$ to be equal to the zeroeth row, with a similar procedure for the other side of the domain.

\section{Numerical Examples} \label{sec:tests}

In the following, we present various numerical tests commonplace in the thermal transport literature.  Our selected tests investigate the optically thin, optically thick, and intermediate regimes.  We therefore exercise each flux formulation described above (i.e., Rusanov, upwinding, and HLL).

In several instances, we report performance diagnostics as measured from serial execution on a \texttt{Xeon Gold 6152} node (\texttt{Skylake})\footnote{\texttt{TTTT} employs \texttt{Python v3.11} and \texttt{NumPy v2.1.3} \citep{harris2020array}.}.  In addition to reporting the spatial zone cycles per second (hereafter, ZCPS) achieved with \texttt{TTTT}, we also report speedups relative to a ``traditional" $S_N$ implementation (using upwind fluxes) in \texttt{Python}\footnote{Rather than running each 
test problem with our traditional $S_N$ \texttt{Python} code, we simply adopt single performance metrics for each configuration studied (vacuum transport: $4.3 \times 10^7$, scattering: $3.2 \times 10^7$, absorption/emission: $3.1 \times 10^7$ angle updates per second).}.  We caveat that we independently evolve both positive and negative $\theta$ angles, which is $2 \times$ the required computational expense in our 2D test problems, and leads to an $2\times$ compression opportunity for our \texttt{TTTT} algorithm.

For all test problems, following \cite{white23}, we select $N_\phi = 2 N_\theta$.  As a simple measure of angular resolution requirements for $S_N$, we expect ray effects to present themselves when the minimum angular extent of a cell at a characteristic length scale $\lambda$ from the source is greater than $\pi / N_\theta = 2 \pi / N_\phi$; therefore, to mitigate ray effects, we require $N_{\theta, \mathrm{crit}} \simeq \pi \lambda / \Delta x$.  For vacuum transport problems, one could take $\lambda$ to be the size of the computational domain $L_x$, thereby giving $N_{\theta, \mathrm{crit}} \simeq \pi N_x$, where $N_x$ is the number of cells resolving the grid in $x$ (assuming $L_x=L_y$ and $\Delta x = \Delta y$).  When radiation interacts with matter, $\lambda$ can be taken to be the mean free path, which leads to the estimate $N_{\theta, \rm crit} \simeq \pi N_x / \bar{\tau}$, where $\bar{\tau} = \max{\left(1, \tau_{\rm min}\right)}$ with $\tau_{\rm min}$ denoting a characteristic minimum optical depth across the simulation domain. We use these estimates to guide our selection of angular resolutions to survey in the numerical examples below.

\subsection{Hohlraum problem} \label{sec:hohlraum}
We first consider the 2D hohlraum problem \citep{mocmc,white23}. This test considers photon propagation from isotropic inner $x$ and inner $y$  boundaries ($I = J = 1$) in vacuum.  The analytic solution for this problem assumes that the isotropic boundaries are of infinite extent; numerically, this can be mitigated by only comparing to the analytics over $x_\mathrm{subset} \in [0, L_x - c t)$, $y_\mathrm{subset} \in [0, L_y - c t)$, such that all rays from the outermost extents of the full domain ($x \in [0, L_x]$, $y \in [0, L_y]$) are causally disconnected from the domain subset at time $t$.  We evolve the system to $t_\mathrm{lim}=0.75$ in a domain of extent $[L_x, L_y]=[2, 2]$ with $c=1$.  Following \cite{white23}, the analytic mean intensity as a function of time is given by
\begin{equation}
    J(t, x, y) = J_\mathrm{left} (t,x,y) + J_\mathrm{bottom} (t,x,y)
\end{equation}
where
\begin{equation}
    J_\mathrm{left} (t,x,y) = \left\{
\begin{array}{cc}
  \frac{1}{2} - \frac{\left(\pi - \eta_x \right) x}{2 \pi c t} - \frac{1}{2 \pi} \sin^{-1} \left( \frac{x \sin \eta_x}{\sqrt{x^2 + y^2}} \right) & \text{ if } x < c t \\
  0 & \phantom{ if } \mathrm{otherwise}
  \end{array}   
\right.  
\end{equation}
\begin{equation}
    J_\mathrm{bottom} (t,x,y) = \left\{
\begin{array}{cc}
  \frac{1}{2} - \frac{\left(\pi - \eta_y \right) y}{2 \pi c t} - \frac{1}{2 \pi} \sin^{-1} \left( \frac{y \sin \eta_y}{\sqrt{x^2 + y^2}} \right) & \text{ if } y < c t \\
  0 & \phantom{ if } \mathrm{otherwise}
  \end{array}   
\right.  
\end{equation} 
and
\begin{equation}
    \eta_x = \cos^{-1} \left[\mathrm{min} \left( \frac{y}{\sqrt{(c t)^2 - x^2}}, 1 \right) \right],
\end{equation}
\begin{equation}
    \eta_y = \cos^{-1} \left[\mathrm{min} \left( \frac{x}{\sqrt{(c t)^2 - y^2}}, 1 \right) \right].
\end{equation}

\begin{figure*}[htb]
    \centering
    \includegraphics[width=\linewidth]{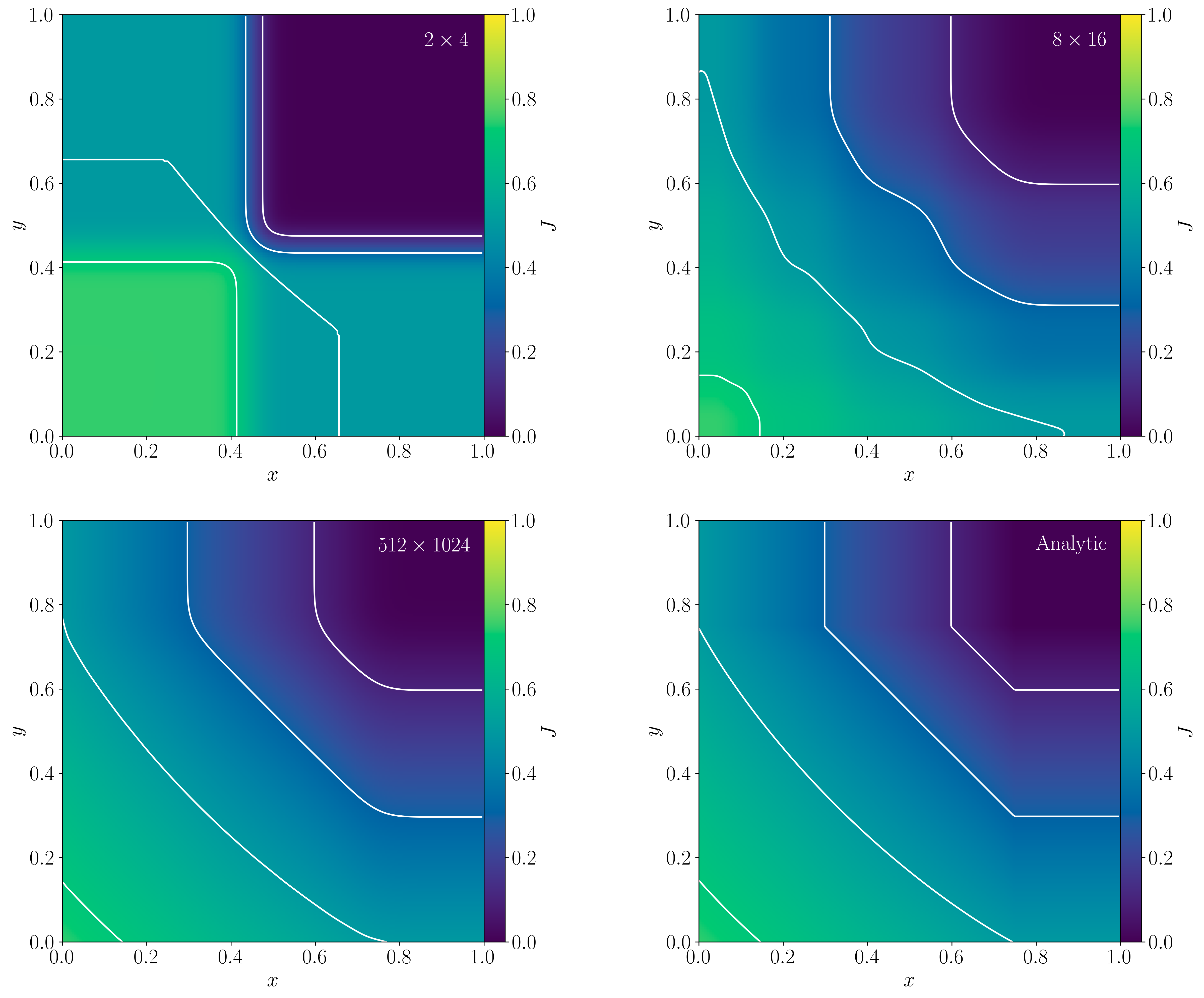}
    \caption{\texttt{TTTT} mean intensity solutions for the 2D hohlraum test obtained using $N_\mathrm{angles} =  2 \times 4 \; \mathrm{(top \; left)}, \; 8 \times 16 \; \mathrm{(top \; right)}, \; 512 \times 1024 \; \mathrm{(bottom \; left)}$. The analytic solution is shown in the bottom right panel.  Each window has the same colorbar range and $J$ contour levels.}
    \label{fig:hohlraum2d}
\end{figure*}

Figure \ref{fig:hohlraum2d} presents solutions at $t_\mathrm{lim} = 0.75$ over $x_\mathrm{subset} \in [0, 1]$, $y_\mathrm{subset} \in [0, 1]$, with $\Delta x = \Delta y = 1/256$, when applying upwind fluxes in \texttt{TTTT} format at three angle counts: $N_\mathrm{angles} = N_\theta \times N_\phi = 2 \times 4 \; \mathrm{(upper \; left)}, \; 8 \times 16 \; \mathrm{(upper \; right)}, \; \mathrm{and} \; 512 \times 1024  \; \mathrm{(bottom \; left)}$ and a rounding tolerance (Frobenius norm) of $\epsilon = 10^{-4}$.  We expect ray effects to be mitigated at $N_{\theta, \mathrm{crit}} \simeq 256\pi$.   The analytic solution is plotted in the bottom right.  Each panel presents contours of $J = [0.1, 0.3, 0.5, 0.7]$.

\begin{figure*}[htb]
    \centering
    \includegraphics[width=\linewidth]{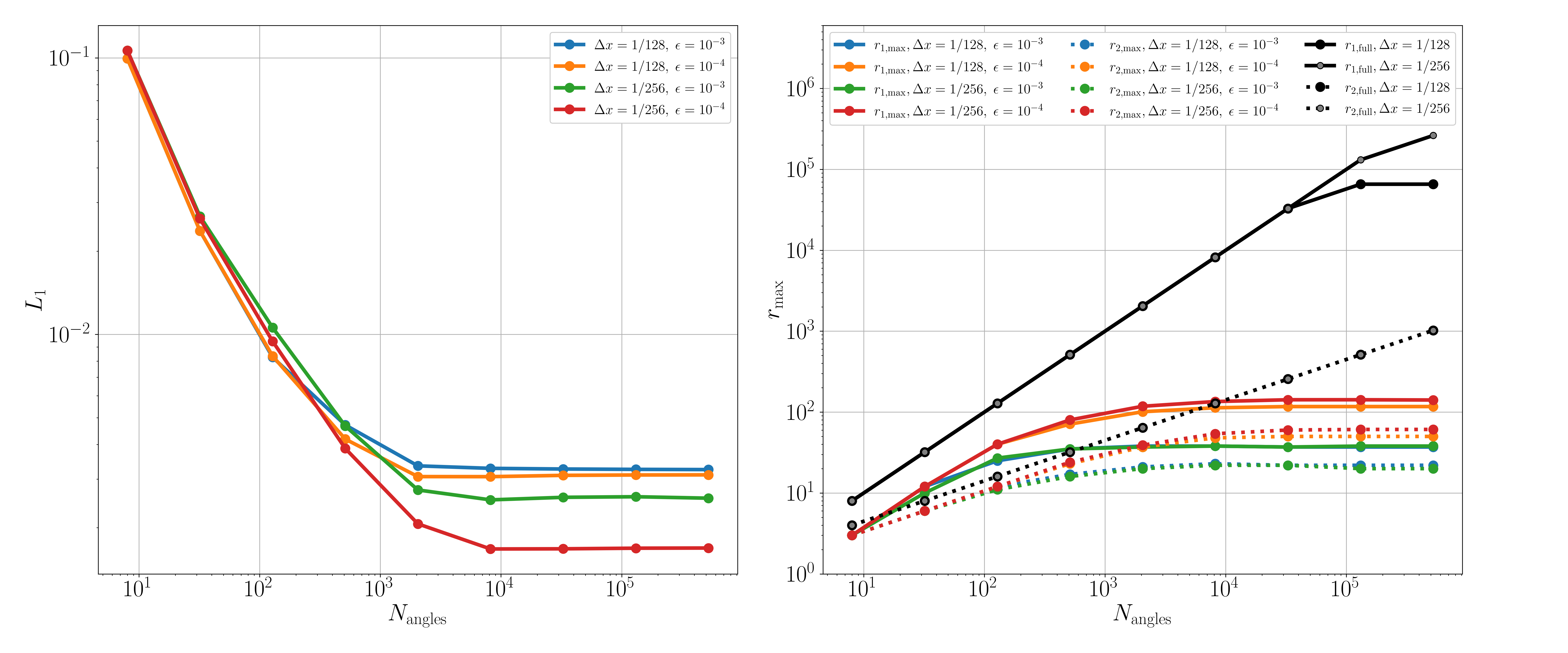}
    \caption{(Left): \texttt{TTTT} mean intensity $L_1$ error convergence with angle for the 2D hohlraum test obtained for rounding tolerances $\epsilon = 10^{-4} \; \mathrm{and} \; = 10^{-3}$ and linear resolutions $\Delta x=1/128 \; \mathrm{and} \; =1/256$.  (Right): Maximum ranks $r_{1,\mathrm{max}}$ and $r_{2,\mathrm{max}}$ achieved in the 2D hohlraum problem as a function of angle for rounding tolerances $\epsilon = 10^{-4} \; \mathrm{and} \; = 10^{-3}$ and linear resolutions $\Delta x=1/128 \; \mathrm{and} \; =1/256$.  ``Full" rank $r_\mathrm{full}$ is shown in black.}
    \label{fig:hohlraum2d_l1}
\end{figure*}

In \texttt{TTTT}, error arises from four sources: (1) temporal discretization error, (2) spatial discretization error, (3) angular discretization error, and (4) error introduced through rounding.  Reducing the error from one source while holding the others fixed leads to significant overall accuracy improvements only while that error remains dominant.  Once that error is subdominant, continued reduction in that error has little overall impact.  This is on display in Figure~\ref{fig:hohlraum2d_l1} (left), which shows the $L_1$ error in the mean intensity as a function of $N_\mathrm{angles}$ for different spatial resolutions and rounding tolerances.  With very few angles, there are severe ray effects that are plainly visible in Figure \ref{fig:hohlraum2d}, suggesting angular discretization is a dominant source of error (at these time steps, spatial resolutions, and rounding tolerances).  Depending on the spatial resolution and rounding tolerance, all four curves flatten to different, but approximately constant, $L_1$ errors, suggesting that angular discretization errors are no longer dominant with a sufficiently large $N_\mathrm{angles}$.  Instead, focusing momentarily on the $\epsilon = 10^{-4}$ cases, we see that reducing $\Delta x$ by a factor of two leads to a concomitant reduction in the asymptotic error by a factor of two, as expected for our spatially first-order method.  However, with $\epsilon = 10^{-3}$, the same reduction in $\Delta x$ does not lead to the expected reduction in asymptotic error, suggesting that now rounding is contributing significantly to the overall error.  This final facet, that a user specified rounding tolerance can play a role in accuracy, is the only aspect of this discussion unique to our \texttt{TTTT} method.

The $N_\mathrm{angles} = 512 \times 1024$ run with $\epsilon = 10^{-4}$ begins with ranks $r_1=r_2 = 1$ and reaches $r_1=141, \; r_2=61$ at $t_\mathrm{lim} = 0.75$.  The \texttt{TTTT} compression factor therefore ranges from $\mathcal{C} = N_x N_y N_\theta N_\phi / \left(r_1 N_x N_y + r_1 r_2 N_\theta + r_2 N_\phi\right) \simeq 5 \times 10^5$ to $\mathcal{C} \simeq 3 \times 10^3$.  Figure \ref{fig:hohlraum2d_l1} (right) presents the extremal ranks $r_\mathrm{1,max}$ and $r_{2, \mathrm{max}}$ achieved for each run surveyed in our convergence study.  We overplot $r_\mathrm{1,full} = \mathrm{min}(N_x N_y, N_\theta N_\phi)$ and $r_{2, \mathrm{full}} = \mathrm{min} (N_x N_y N_\theta, N_\phi)$ which represent ``full rank" (i.e., the upper bounds for ranks for the \texttt{TTTT} calculation).  We will examine the performance history for a vacuum transport problem in \S\ref{sec:ls2d}.

\subsection{Thermal relaxation}
\label{sec:relax}
We next test the operator split, backward Euler, absorption/emission source term presented in \S \ref{sec:rhs} via a thermal relaxation test problem \citep[c.f.,][]{jiang14, jiang21}. We initialize each of 4 spatial cells in a 1D grid (with $\Delta x=1$) with material density $\rho = 1$ and temperature $T = 2$ everywhere.  We initialize the radiation field with $I = J = c E / (4\pi)$ everywhere, where $E = a_r T_r^4$, and the radiation temperature $T_r = 1$.  We set $c_v = 8$, $c = 1$, and apply periodic boundary conditions.  

\begin{figure*}[htb]
    \centering
    \includegraphics[width=\linewidth]{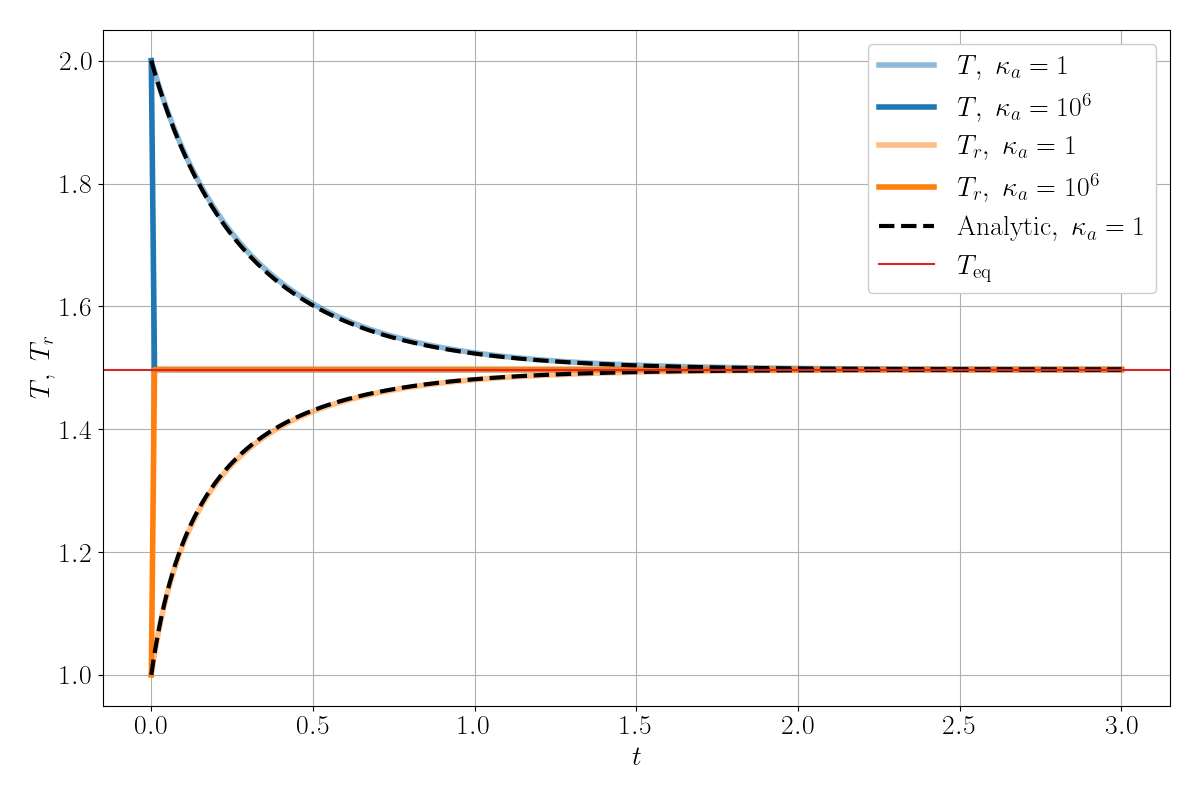}
    \caption{\texttt{TTTT} thermal relaxation test with varying specific absorption opacities $\kappa= 1$ (translucent) and $\kappa = 10^6$ (opaque).  The analytic temporal evolution is shown for the $\kappa= 1$ in dashed black.  The analytic equilibrium temperature is shown in red. 
 The $\kappa = 10^6$ models obtains thermal equilibrium within a single timestep.}
    \label{fig:relax}
\end{figure*}

This setup initializes the material temperature and radiation temperature out of equilibrium; the radiation source term will drive them to the equilibrium temperature defined via energy conservation
\begin{equation}
    \rho c_v T_i + a_r T_{r,i}^4 = \rho c_v T_\mathrm{eq} + a_r T_\mathrm{eq}^4.
\end{equation}
We set $\kappa_a \equiv \mathrm{constant}$. The magnitude of $\kappa_a$ determines the relaxation timescale $t_a = [c \rho \kappa_a]^{-1}$.  We test $\kappa_a = 1$ and $10^{6}$ and evolve the system to $t_\mathrm{lim} = 3$.  Figure \ref{fig:relax} presents the numerical solutions for each test (material temperature: blue, radiation temperature: orange).  For the $\kappa_a = 10^6$ run (opaque), the problem is incredibly stiff, and the \texttt{TTTT} solver achieves thermal equilibrium within a single backward Euler timestep. For the $\kappa_a = 1$ run, we can recover the analytic temporal evolution (dashed black) when using 300 timesteps to evolve to $t_\mathrm{lim}=3$.  When using the light crossing time for the numerical timestep, first order temporal errors due to our backward Euler discretization overestimate the thermalization time.  These tests invoke upwind fluxes and $N=512 \times 1024$ angles, however, neither angular discretization nor flux selection affect the solutions, as isotropy is maintained throughout, thereby promoting a zero flux divergence.  This low rank structure is reflected via ranks $r_1 = 1, \; r_2 = 1$ maintained throughout the entire evolution for the $\kappa = 1$ and $= 10^{6}$ runs, corresponding to $\mathcal{C} \simeq 1.4 \times 10^3$. A detailed look at rank and performance in the very optically thick regime will be examined in \S \ref{sec:gauss2d}.

\subsection{Gaussian diffusion} \label{sec:gauss2d}
Next we consider the diffusion of a 2D Gaussian pulse due to isotropic scattering, with analytic solution
\begin{equation}
    E(t,x,y) = \frac{A}{ 4 \pi D (t_0 + t)} \exp \left[- \frac{x^2 + y^2}{4 D (t_0 + t)} \right],
\end{equation}
where $t_0$ characterizes an ``initial time" setting the initial width of the Gaussian, $D = c / (3 \rho \kappa_s)$ is the scattering diffusion coefficient, and $A$ is an amplitude \citep{sekora10, jiang14, jiang21}.

\begin{figure*}[htb]
    \centering
    \includegraphics[width=\linewidth]{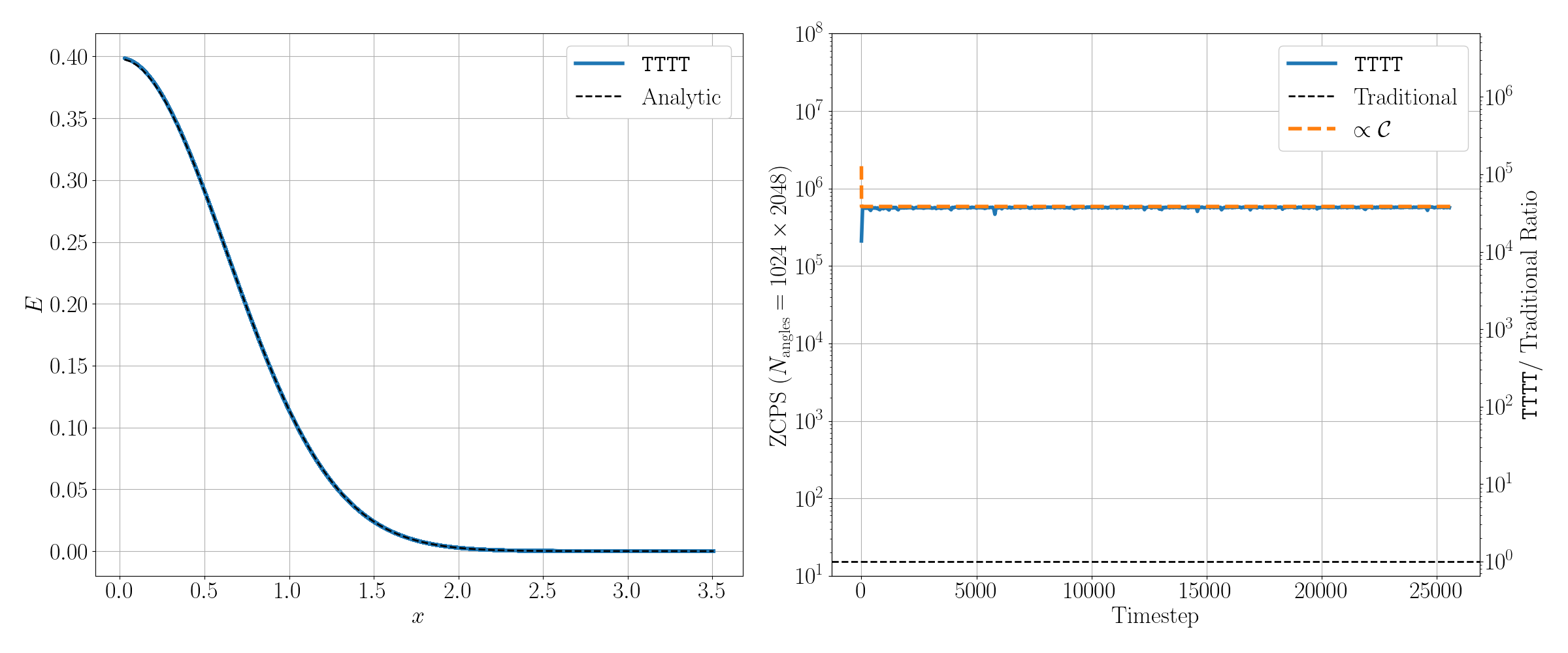}
    \caption{(Left): \texttt{TTTT} radiation energy density solution for the 2D Gaussian diffusion test obtained using $N_\mathrm{angles} = 1024 \times 2048$. Analytics are overplotted in dashed black. 
 (Right): \texttt{TTTT} 2D Gaussian diffusion serial performance (in zone cycles per second, ZCPS).  Speedups relative to traditional methods are reported assuming a reference metric of $3.2 \times 10^{7}$ angle updates per second.}
    \label{fig:gaussian2d}
\end{figure*}

We initialize the radiation field via $I(x,y,z) = J(x,y,z) = c E(0,x,y,z) / (4\pi)$ and apply outflow boundary conditions on a domain of extent $[L_x, L_y] = [5, 5]$ resolved by $128^2$ cells.  Given a large enough optical depth per cell $\tau_c = \rho \kappa_s \Delta x$, the upwind flux struggles to recover analytics.  We here apply the Rusanov flux, selecting $\rho \kappa_s = 10^3$ such that we can stably integrate the problem with $S^{+} = 0$. 
Figure \ref{fig:gaussian2d} (left) demonstrates recovery of analytics when evolving the system to $t_\mathrm{lim} = t_0 + 400$, with $\rho = 1$, $A = 1$, $t_0 = 200$, and $c = 1$.

For this calculation, we apply $N_\mathrm{angles}=1024 \times 2048$. The optical depth per cell in this problem is $\tau_c \simeq 40$ and therefore $N_{\theta,\mathrm{crit}} \simeq \pi N_x / \bar{\tau} \simeq \pi/\tau_c \ll 1$.  As such, the angle count applied is absurdly large for this problem.  Put differently, the large scattering opacity demands that the radiation field maintain isotropy and hence low rank structure throughout the duration of the simulation.  With these caveats, we here examine the rank structure and performance history for this problem as a ``best-case scenario" for \texttt{TTTT} (i.e., the optically thick regime).  The ranks remain constant in this problem at $[r_1, r_2] = [3, 3]$ corresponding to a compression factor $\mathcal{C} \simeq 5.3 \times 10^{5}$.  Figure \ref{fig:gaussian2d} (right) presents the (spatial) zone cycles per second achieved as a function of timestep for the 2D Gaussian problem (blue).  The black line corresponds to our traditional $S_N$ metric for performance comparison.  The orange curve is $\propto \mathcal{C}$, motivated by the idea that speedups associated with tensors (in the absence of overhead---e.g., rounding) should be proportional to the compression ratio.  We find that tensors achieve $\sim 6 \times 10^{5}$ ZCPS ($\simeq 10^{12}$ angle updates per second) amounting to a $\sim 4 \times 10^{4}$ speedup relative to traditional $S_N$ implementations.  Ranks $r_1, r_2$ for this problem remain quite constant throughout the duration of the calculation as shown by the curve $\propto \mathcal{C}$ (orange) and reflected in the raw \texttt{TTTT} performance measure (blue).  In the next test (\S \ref{sec:ls2d}), we examine the rank structure and performance in the entirely opposite regime where ranks significantly inflate in time.

\subsection{Line source} \label{sec:ls2d}
The 2D line source problem \citep[see, e.g.,][]{ls2d01, garrett13, geo_ray23} considers an isotropic pulse of radiation where the radiation energy density is initialized via a delta function
\begin{equation}
    E(t=0, x, y) = \mathcal{U}_0 \; \delta(x, y).
\end{equation}
Following \cite{geo_ray23}, the analytic solution to the evolution of this pulse in vacuum is given by
\begin{equation}
  E(t,x,y) = \frac{\mathcal{U}_0}{2 \pi} \frac{\mathbf{H} [(c t - \sqrt{x^2 + y^2})]}{c t \sqrt{(c t)^2 - (x^2 + y^2)}} 
\end{equation}
where $\mathbf{H[]}$ is the Heaviside step function. Because we cannot initialize a delta function on our discrete spatial grid, we invoke a Gaussian
 \begin{equation}
     \tilde{E}(t=0, x, y) = \frac{\mathcal{U}_0}{2 \pi \omega^2} \exp \left[-\frac{x^2 + y^2}{2 \omega^2} \right],
 \end{equation}
with a width specified by parameter $\omega$. The analytic solution to the evolution of this Gaussian pulse requires taking a convolution
\begin{equation}
    \tilde{E} (t, x, y) = \iint_{\mathbb{R}^2} E(t, x- x^\prime, y - y^\prime) \tilde{E} (t = 0, x^\prime, y^\prime) dx^\prime dy^\prime,
\end{equation}
that ensures $\iint_{\mathbb{R}^2} E(t,x,y) dx dy = \iint_{\mathbb{R}^2} \tilde{E}(t,x,y) dx dy = \mathcal{U}_0$.

This problem is specifically designed to demonstrate ray effects in $S_N$-like schemes.  The pulse is initially isotropic, with an expected low rank structure, but as the problem progresses, an axisymmetric ring expands outward at $c$, requiring more and more angles to recover the analytic solution.  At a final time $t_\mathrm{lim}$, the radius of the axisymmetric ring is $R_\mathrm{lim} \sim c t_\mathrm{lim}$, hence requiring $N_{\phi,\mathrm{crit}} \simeq 2 \pi R_\mathrm{lim} / \Delta x$ to mitigate ray effects.  Put differently, for any fixed $\Delta x$ and $N_\mathrm{angles}$, one could always find a final time (and hence box size) that presents ray effects and/or high rank structure.  This then makes this problem an opposite regime to the one presented in \S \ref{sec:gauss2d}, representing a near ``worst-case scenario" for \texttt{TTTT}.  

We resolve a square grid ($[-1.1, 1.1]$ edges) with $500^2$ spatial cells, selecting $c=1$, $\omega=0.03$, $\mathcal{U}_0 = 1$, and $\mathrm{CFL} \simeq 0.4$, and initializing the specific intensity via $I(x,y) = J(x, y) = c \tilde{E}(t=0, x, y) / (4 \pi)$. Compared to previously published versions of this test, our $\Delta x$ is slightly smaller to mitigate spatial error observed at extremely high angle counts.  

\begin{figure*}[htb]
    \centering
    \includegraphics[width=\linewidth]{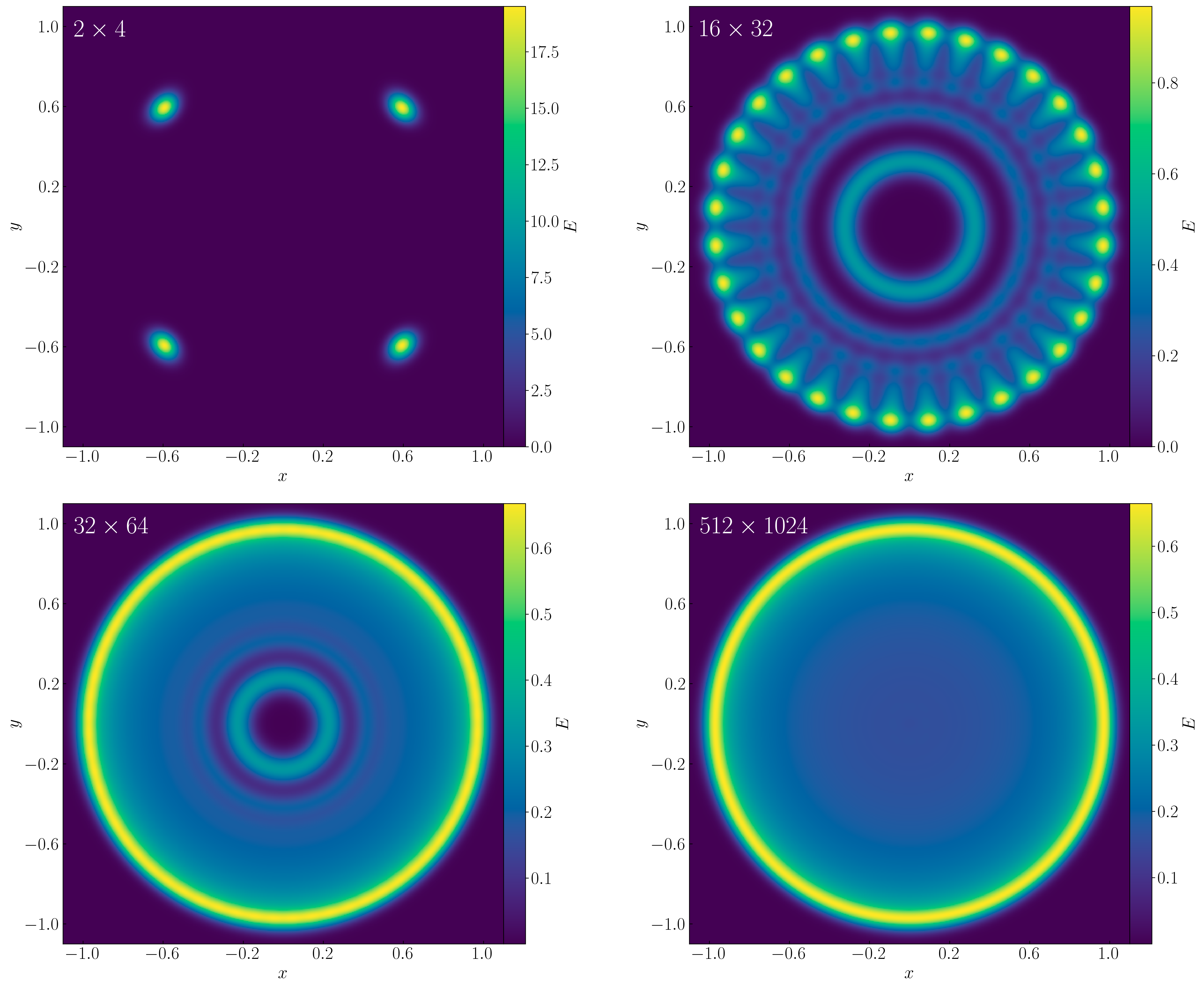}
    \caption{Solutions to the line source problem obtained using $N_\mathrm{angles} =  2 \times 4 \; \mathrm{(top \; left)}, \; 16 \times 32 \; \mathrm{(top \; right)}, \; 32\times 64 \; \mathrm{(bottom \; left)}, \; \mathrm{and} \; 512 \times 1024 \; \mathrm{(bottom\;right)}$.  Note the change in the colorbar limits for each panel.}
    \label{fig:ls2d}
\end{figure*}

\begin{figure*}[htb]
    \centering
    \includegraphics[width=\linewidth]{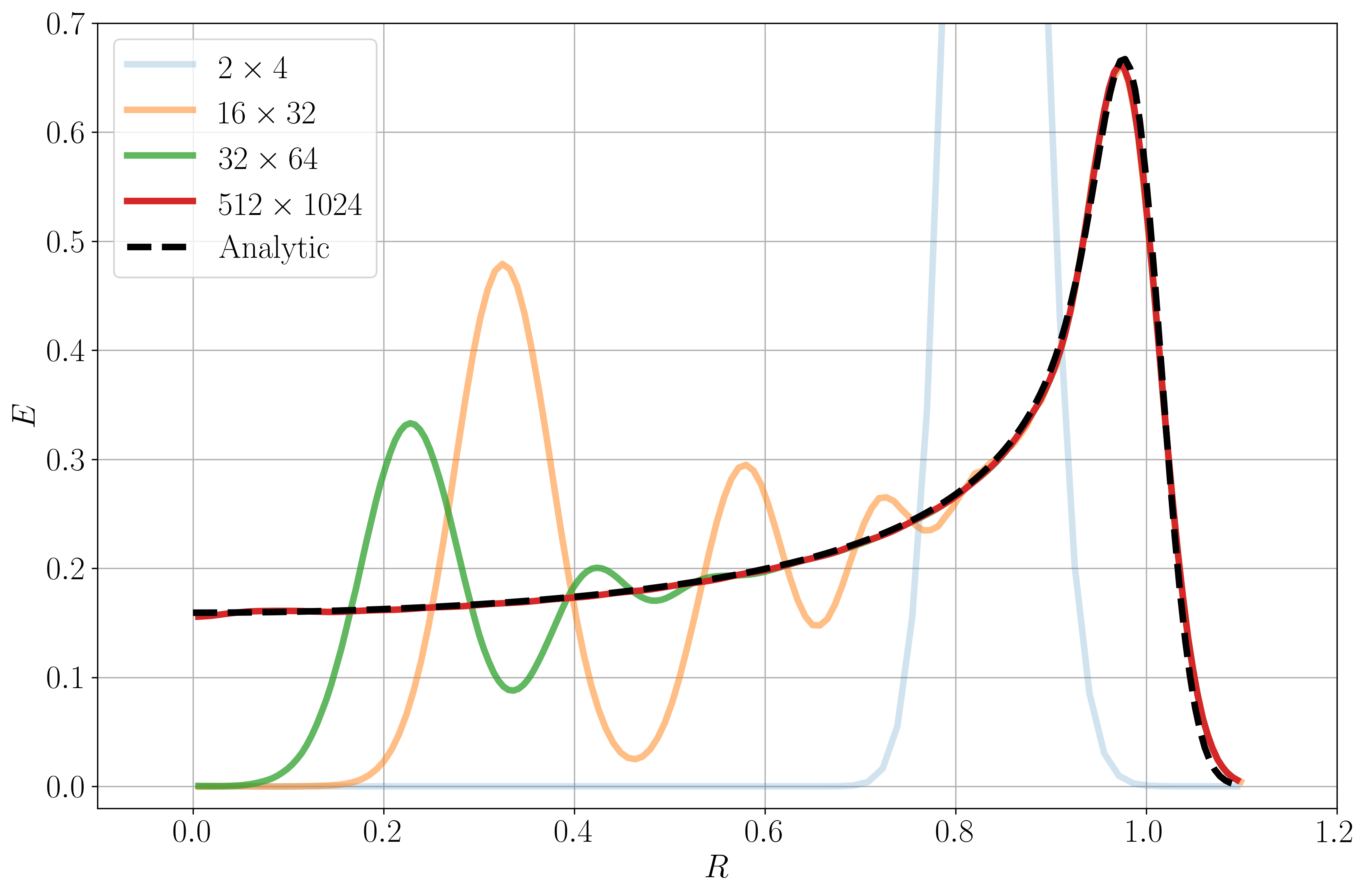}
    \caption{Azimuthally averaged line source solutions as a function of cylindrical radius for $N_\mathrm{angles} = 2 \times 4 \; \mathrm{(blue)}, 16 \times 32 \; \mathrm{(orange)}, 32 \times 64 \; \mathrm{(green)}, \; \mathrm{and} \; 512 \times 1024 \; \mathrm{(red)}$.  The analytic solution is presented in dashed black.}
    \label{fig:ls2d_avg}
\end{figure*}

Figure \ref{fig:ls2d} presents solutions at $t_\mathrm{lim} = 1$ with a rounding tolerance $\epsilon = 10^{-4}$ at four angle counts: $N_\mathrm{angles} = 2 \times 4, \; 16 \times 32, \; 32 \times 64, \; \mathrm{and} \; 512 \times 1024$. The $N_\mathrm{angles} = 2 \times 4$ solution shows 4 discrete pulses (corresponding to the 8 angles with matching $\lvert \theta_\ell \rvert$).  At $N_\mathrm{angles} = 32 \times 64$, the $R_\mathrm{lim}$ axisymmetric ring is beginning to be resolved, but a hole in the center of the domain is observed due to a lack of angles near the poles of the angular mesh with radius $R_\mathrm{min} \simeq c t \sin \theta_\mathrm{min}$, where $\theta_\mathrm{min}$ is the minimum $\theta$ angle represented via the angular mesh.  At $N_\mathrm{angles} = 512 \times 1024$, ray effects have nearly vanished.  To demonstrate recovery of analytics, Figure \ref{fig:ls2d_avg} presents azimuthally averaged solutions compared to the exact solution. Figure \ref{fig:ls2d_avg} demonstrates that the most slowly converging error is that associated with a lack of angles near the poles of the angular mesh.  To mitigate this discretization error, we require $R_\mathrm{min} < \Delta x$, which is a stringent requirement exacerbated via our latitude-longitude quadrature.  Other quadrature choices might improve the $R_\mathrm{min}$ errors, but we reinforce that our TT decomposition requires an angular discretization expressible as a 2D tensor product grid.  The $N_\mathrm{angles} = 512 \times 1024$ solution recovers the exact solution quite nicely, with only slight deviations from analytics associated with (1) spatial and/or temporal errors and (2) $R_\mathrm{min}$ errors. 
 
\begin{figure*}[htb]
    \centering
    \includegraphics[width=\linewidth]{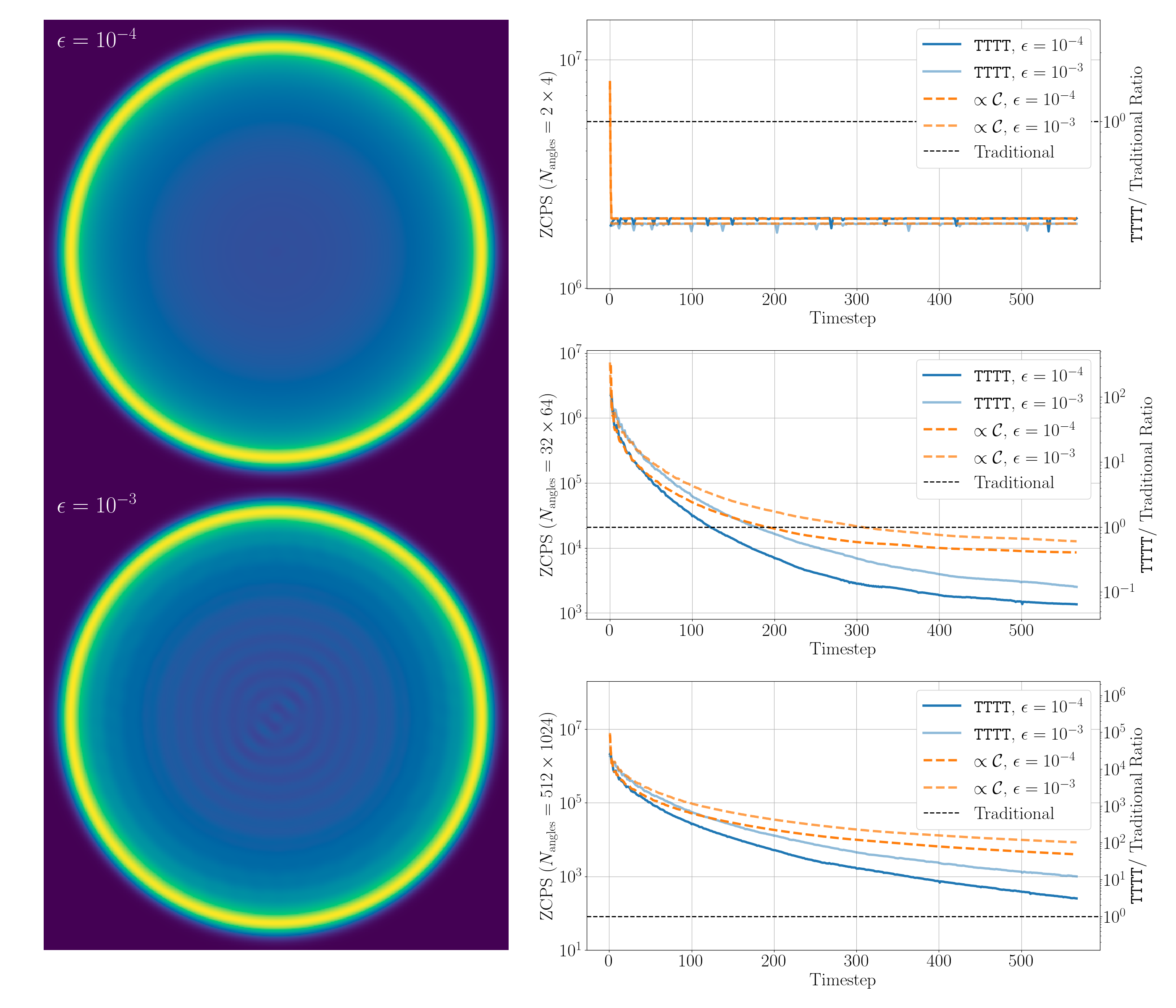}
    \caption{(Left): \texttt{TTTT} line source solutions for $N_\mathrm{angles} = 512  \times 1024$ for two tolerances thresholds: $10^{-4}$ (top) and $10^{-3}$ (bottom).  (Right): \texttt{TTTT} line source serial performance (in zone cycles per second, ZCPS) for the two tolerances at $N_\mathrm{angles} = 2  \times 4 \; \mathrm{(top)}, 32  \times 64 \; \mathrm{(middle)},  \; \mathrm{and} \; 512  \times 1024 \; \mathrm{(bottom)}$.  Speedups relative to traditional methods are reported assuming a reference of $4.3 \times 10^{7}$ angle updates per second---the $N_\mathrm{angle} = 512\times1024$ problem cannot actually be run serially without tensor compression on our \texttt{Skylake} testbed.}
    \label{fig:ls2d_perf}
\end{figure*}

To investigate the effects of more aggressive rounding, we also consider behavior with a rounding tolerance $\epsilon = 10^{-3}$ in Figure \ref{fig:ls2d_perf} (left).  Numerical ringing is prevalent in the $\epsilon = 10^{-3}$ solution.  The integrated radiation energy density has a conservation error of $2 \times 10^{-4} \; \mathcal{U}_0$, whereas the $\epsilon = 10^{-4}$ presents a $8 \times 10^{-5} \; \mathcal{U}_0$ error.  More aggressive rounding does lead to more performant integrations (i.e., lower ranks).  Figure \ref{fig:ls2d_perf} (right) presents the performance history (blue) of the line source problem for $N_\mathrm{angles} = 2 \times 4 \; \mathrm{(top)}, \; 32 \times 64 \; \mathrm{(middle)}, \; \mathrm{and} \; 512 \times 1024 \; \mathrm{(bottom)}$ with rounding tolerances $\epsilon = 10^{-4} \; \mathrm{(opaque)} \; \mathrm{and} \; 10^{-3} \; \mathrm{(translucent)}$. We overplot our traditional $S_N$ performance metric (black) and overplot a curve $\propto \mathcal{C}$ (orange). In the $N_\mathrm{angles} = 2\times 4$ calculations, both rounding tolerances present ranks near ``full" rank, thereby leading to flat performance curves.  In the $N_\mathrm{angles} = 32 \times 64$ runs, early evolution shows almost a factor of 100$\times$ speedup relative to traditional $S_N$, but as the ranks of the calculations continue to increase, \texttt{TTTT} eventually becomes less performant than traditional methods (with e.g., $r_1 = 167, \; r_2 = 39$ at timestep 120 for the $\epsilon = 10^{-4}$ run).  For the same run with $\epsilon = 10^{-3}$, performance is enhanced by $\sim 3 \times$. Both runs exhibit deviations from curves $\propto \mathcal{C}$, hinting at significant overhead associated with rounding operations.  The $N_\mathrm{angles} = 512 \times 1024$ solution shows speedups up to $\sim 10^{4}$ in early evolution but drops to only $12 \times$ and $3 \times$ speedups at the final timestep for $\epsilon = 10^{-3}$ and $\epsilon = 10^{-4}$, respectively.  At our final snapshots, we find compression factors of $\mathcal{C} \simeq 2, 3, \; \mathrm{and} \; 290$ ($\epsilon=10^{-4}$) and $\mathcal{C} \simeq 2, 4, \; \mathrm{and} \; 560$ ($\epsilon=10^{-3}$) for the $N_\mathrm{angles} = 2 \times 4, \; 32 \times 64, \; \mathrm{and} \; 512\times1024$ runs, respectively.  Running at $N_\mathrm{angles} = 512\times1024$ with traditional methods would require $\simeq1$TB of memory to store the specific intensity solution vector, making the problem prohibitively memory expensive for our single \texttt{Skylake} node testbed. 

We emphasize that for a fixed $N_\mathrm{angle}$ and $\Delta x$, we could always extend the problem duration $t_\mathrm{lim}$ such that the simulation always approaches high rank (wherein the \texttt{TTTT} performance curves flatten).

\subsection{Crooked Pipe} \label{sec:crooked}
The 2D Gaussian test exhibits low rank structure in TT format; the line source test exhibits rank growth as the simulation progresses, with a few examples of \texttt{TTTT} performance lagging that of a traditional $S_N$ implementation. To this end, we now move our attention to a test which contains both regimes simultaneously: the crooked pipe test problem.  Following the setup prescribed in \cite{southworth24} (see their Figure 7), this problem invokes an isotropic 0.5 keV boundary source incident on a crooked pipe.  Walls of the pipe are optically thick; the interior is optically thin.  As the problem evolves, radiation propagates through the pipe, before reaching an obstacle in the complicated geometry.  The walls heat and themselves radiate, photons propagate around the obstacle and, given enough time, emerge from the other side of the crooked pipe.

Our computational domain spans $x \in [-3.5 \; \mathrm{cm}, 3.5 \; \mathrm{cm}]$ and $y \in [-2.0 \; \mathrm{cm}, 2.0 \; \mathrm{cm}]$. The walls have density $\rho_\mathrm{wall} = 10 \; \mathrm{g/cm^{3}}$, specific absorption opacity $\kappa_{a, \mathrm{wall}} = 200 \; \mathrm{cm^{2}/g}$, and specific heat at constant volume $c_{v, \mathrm{wall}} = 10^{15} \; \mathrm{erg/g/keV}$.  Interior to the pipe we have density $\rho_\mathrm{pipe} = 0.01 \; \mathrm{g/cm^{3}}$, specific absorption opacity $\kappa_{a, \mathrm{pipe}} = 20 \; \mathrm{cm^{2}/g}$, and specific heat at constant volume $c_{v, \mathrm{pipe}} = c_{v, \mathrm{wall}}$.  The specific scattering opacity $\kappa_s$ is everywhere zero.  Everywhere in the active domain, a background material temperature is set such that it is in thermal equilibrium with an everywhere isotropic background radiation field wherein $T_{\mathrm{background}} = T_{r, \mathrm{background}} = 50 \; \mathrm{eV}$.  The radiation field in the active domain is therefore set via $I = J = c [a_r T_{r, \mathrm{background}}^4] / (4 \pi)$.  Boundary conditions are everywhere outflow except for the inner-$x$ radiation boundary condition, where we setup an isotropic boundary source over $-0.5 \mathrm{\;cm} \leq y \leq 0.5 \mathrm{\;cm}$ where $T_\mathrm{source} = T_{r, \mathrm{source}}$ and $I = J = c [a_r T_{r, \mathrm{source}}^4] / (4 \pi)$ with $T_{r, \mathrm{source}} = 0.5 \; \mathrm{keV}$. The problem is run to a final time $t_f = 100 \; \mathrm{ns}$.  We invoke $\mathrm{CFL} \simeq 0.4$ and the HLL flux with $\beta \simeq 5$.

\begin{figure*}[htb]
    \centering
    \includegraphics[width=\linewidth]{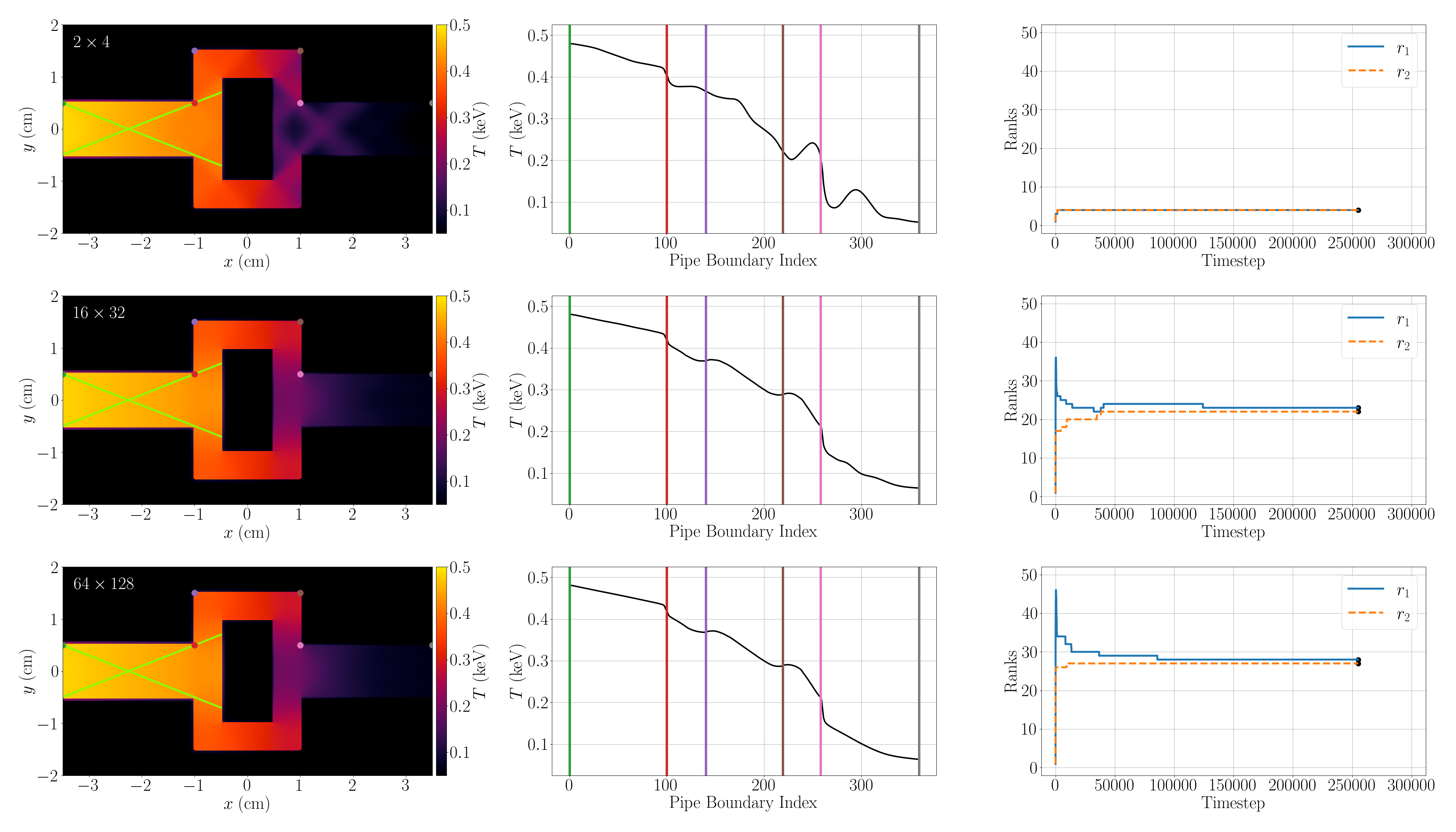}
    \caption{Crooked pipe solutions at fixed linear resolution but varying angular resolution.  The top, middle, and bottom rows correspond to $N_\mathrm{angles} = 2 \times 4$, $16 \times 32$, and $64 \times 128$.  The left column shows the material temperature across the domain for each angular resolution at $t \simeq 85$ ns; lime green lines show exact boundaries for shadows cast by the pipe boundaries during the initial transient; colored markers note positions indicated via vertical bars in the middle column.  The middle column shows line outs of the material temperature along the pipe top boundary.  The right column shows the tensor ranks history for the calculation as a function of iteration index.  The associated movie shows the time evolution of the problem over 100 ns simulation time. The movie duration is 10 seconds.}
    \label{fig:crooked}
\end{figure*}

\begin{figure*}[htb]
    \centering
    \includegraphics[width=0.65\linewidth]{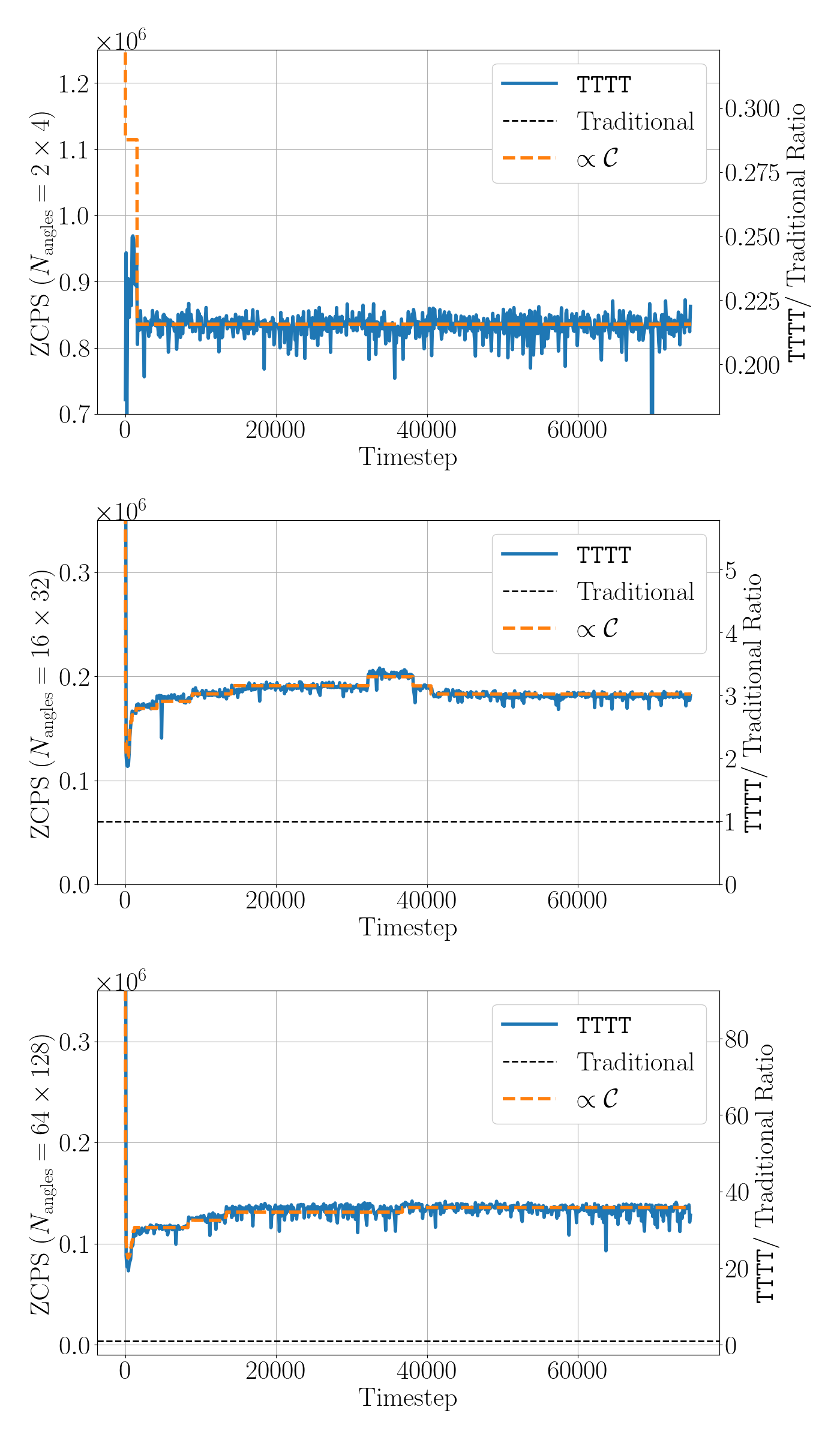}
    \caption{\texttt{TTTT} crooked pipe serial performance (in zone cycles per second, ZCPS) for the first 75,000 explicit timesteps of the run using $N_\mathrm{angles} = 2  \times 4 \; \mathrm{(top)}, \; 16  \times 32 \; \mathrm{(middle)},  \; \mathrm{and} \; 64  \times 128 \; \mathrm{(bottom)}$.  Speedups relative to traditional methods are reported assuming a reference of $3.1 \times 10^{7}$ angle updates per second.}
    \label{fig:crooked_perf}
\end{figure*}

Figure \ref{fig:crooked} (left) presents numerical solutions of the gas temperature obtained using $N_\mathrm{angles} = 2 \times 4 \; (\mathrm{top}), \; 16 \times 32  \; (\mathrm{middle}), \; \mathrm{and} \; 64 \times 128  \; (\mathrm{bottom})$ when invoking an $[N_x, N_y] = [280, 160]$ grid. Figure \ref{fig:crooked} (middle) panels correspond to lineouts along the pipe upper wall, with colored vertical bars corresponding to locations denoted by markers in the left column panels.  Figure \ref{fig:crooked} (right) panels track the rank history of the problem.  Figure \ref{fig:crooked_perf} reports the performance history for all three angle counts.  

As the simulation begins, photons from the left boundary source propagate through the optically thin pipe before reaching the obstacle.  Many angles are required in this transient part of the problem, with only the high angle count runs recovering the expected shadow boundaries traced by the green lines in Figure \ref{fig:crooked} (left).  The $N_\mathrm{angle} = 2 \times 4$ run has no angle normals $\mathbf{n}$ directed enough in the $\hat{x}$ direction such that light emitted from the source reaches the first obstacle within a light crossing time. Moreover, this low angle run produces local ``hot spots" on pipe boundaries associated with energy directed into few ordinates striking the pipe boundaries at fixed spatial locations.  The high angle count runs heat the pipe walls smoothly (see the lineouts in Figure \ref{fig:crooked}).  As the simulation progresses and pipe wall heating/emission continues, the radiation field isotropizes leading to lower rank structure, with angular structure in the radiation field most important near the radiation front propagating around the obstacle.  Eventually, the rank structure of the calculation stabilizes for the remainder of the simulation.  Herein, \texttt{TTTT} supplies $\sim20 \times$ and $\sim 275 \times$ compressions $\sim 3 \times$ and $\sim 34\times$ speedups relative to traditional $S_N$ methods for the $N_\mathrm{angle} = 16 \times 32$ and $N_\mathrm{angle} = 64 \times 128$ runs, respectively\footnote{We expect these speedup measures to be conservative, as \texttt{TTTT} applies the \cite{jiang21} HLL flux, whereas the traditional $S_N$ performance metric invoked upwinding.}. Interestingly, even the $N_\mathrm{angle} = 16 \times 32$ and $N_\mathrm{angle} = 64 \times 128$ calculations present qualitative differences (e.g., sharp features in the lower resolution lineout not visible in the smooth high angle counterpart), pointing at lack of angular convergence in the intermediate resolution run.  Taking $\lambda = 1 \; \mathrm{cm}$, we find $N_{\theta,\mathrm{crit}} \simeq \pi \lambda/\Delta x \simeq 128$, implying the possible need for even large $N_\mathrm{angle}$ for fully angularly converged solutions.  

\subsection{Stellar Irradiation} \label{sec:stellar}
Finally, we consider an astrophysically relevant test: a 2D variant of the stellar irradiation problem of \cite{klassen14}.  In this test, two stars irradiate a central dense clump.  Such setups are used as precursors for radiation hydrodynamic simulations of clump photoevaporation \citep{rijkhorst06}.  Previous iterations of this test invoke ray tracing to find equilibrium solutions; here, we seek to demonstrate that we can run a similar problem \textit{with $S_N$} when throwing many angles at the problem with \texttt{TTTT}.  Albeit the \cite{klassen14} and \cite{rijkhorst06} runs are performed in 3D, our target demonstration only requires 2D.  As such, parameters in our models are slightly modified relative to earlier works.

An ambient background with $\rho_b = 10^{-20} \; \mathrm{g/cm^3}$ and $T_b = T_{r,b} = 1 \; \mathrm{K}$ permeates a square computational domain with $x$- and $y$- domain boundaries $[-1000,1000] \; \mathrm{AU}$.  We impose a constant specific absorption opacity $\kappa_a = 100 \; \mathrm{cm^2 / g}$ everywhere (we do not introduce scattering in this test). The specific heat at constant volume is everywhere $c_v = 1.2472 \times 10^{8} \; \mathrm{erg/g/K}$. Centered on the domain origin, a cold clump ($\rho_c = 10^{-16} \; \mathrm{g/cm^3}$ and $T_c = T_{r,c} = 1 \; \mathrm{K}$) is initialized with radius $R_c = 267 \; \mathrm{AU}$.  Centered on $[0, -500] \; \mathrm{AU}$ and $[-500, 0] \; \mathrm{AU}$, we model two stars.  The stellar radii are below the grid scale, therefore, we mock-up their radiation fields by imposing gas and radiation temperatures $T_s = T_{r,s} = 140 \; \mathrm{K}$  within radii $R_s = 8 \; \mathrm{AU}$ (hence giving $c \pi R_s^2 a_r T_s^4 = L_\odot$). We artificially lock $T_s = T_{r,s} = 140 \; \mathrm{K}$ within $R_s$ throughout the course of the simulation by imposing density $\rho_s = 10^{-12} \; \mathrm{g} / \mathrm{cm^3}$.

We resolve a grid by $500^2$ cells and invoke the Rusanov flux with $S^{+} = c$, imposing outflow boundary conditions everywhere.  We apply a rounding tolerance $\epsilon = 10^{-4}$.

\begin{figure*}[htb]
    \centering
    \includegraphics[width=\linewidth]{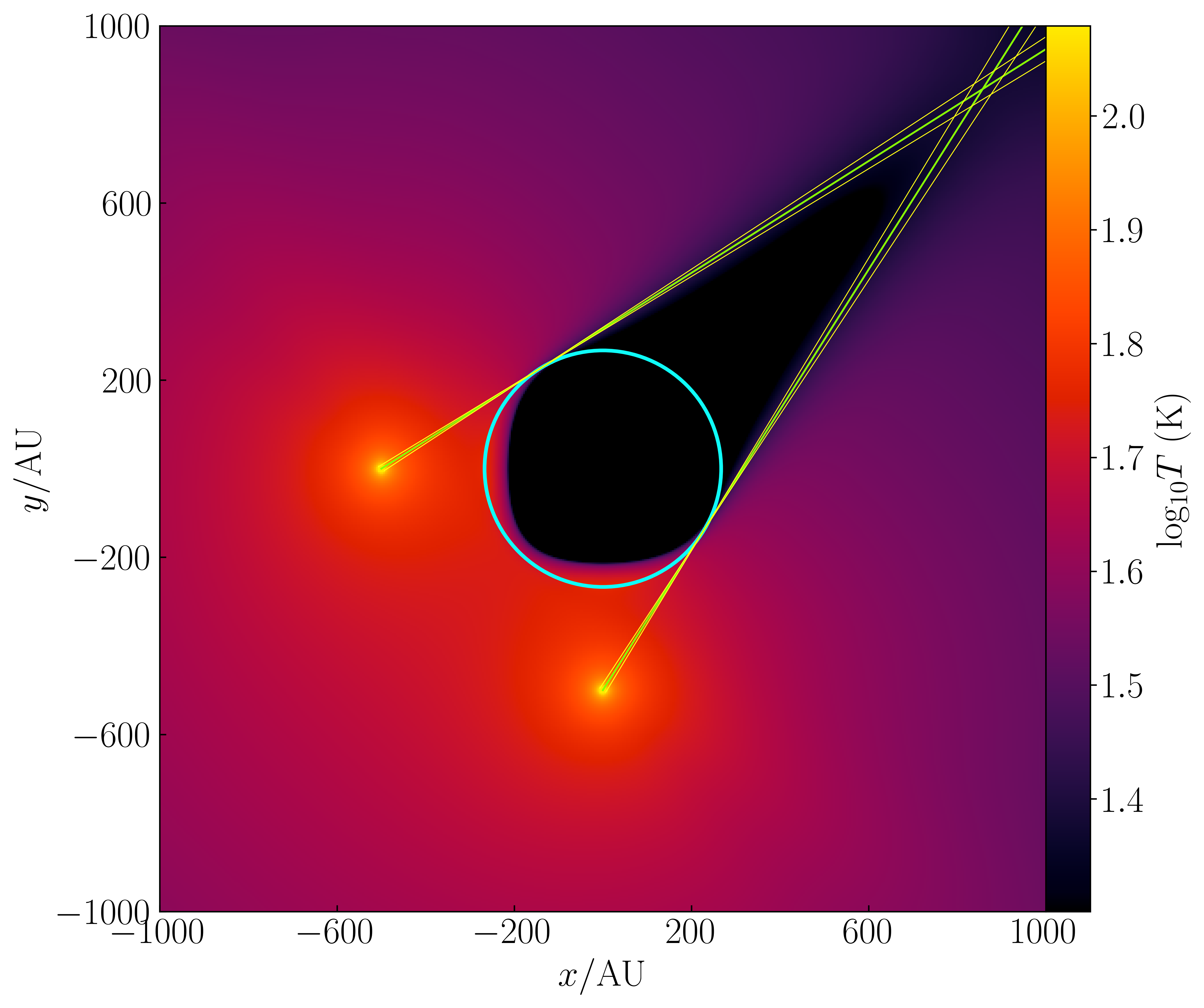}
    \caption{Irradiation of an overdense clump by two stars.  The blue circle shows clump boundaries. Green lines inform predicted shadow regions by tracing geodesics emanating from the stellar sources and tangent to the clump. Due to the finite size of our stellar sources, yellow lines track geodesics emanating from source boundaries.  This calculation uses $N_\mathrm{angles} = 512 \times 1024$ and cannot actually be run serially without tensor compression on our \texttt{Skylake} testbed.}
    \label{fig:stellar} 
\end{figure*}

Figure \ref{fig:stellar} presents material temperature solutions obtained using $N_\mathrm{angles} = 512 \times 1024$ angles ($N_{\theta,\mathrm{crit}} \simeq 500 \pi$). Shadows emerge behind the clump.  Green lines trace exact shadow boundaries.  Because our stellar sources are of finite size, we additionally overplot yellow lines which inform penumbral regions.  As the solution approaches steady state, it exhibits ranks $r_1 = 424, \; r_2 = 84$, corresponding to a factor of $\simeq 10^3 \times$ compression.  Put differently, for this problem, \texttt{TTTT} can represent the specific intensity solution vector with only $\simeq$1 GB, compared to a $\simeq$1 TB storage requirement for traditional $S_N$ (again, a nonviable memory demand for our single \texttt{Skylake} node). Even if we could fit this problem on this compute node, serial execution gives (under the bigger assumption of continued $3.1 \times 10^7$ angle updates per second for traditional $S_N$) an integrated time to solution of $\simeq$78 days; \texttt{TTTT} arrives at the solution in $\simeq$30 hr (a $\simeq$60$\times$ speedup). 

\section{Discussion} \label{sec:discussion}
The viability of \texttt{TTTT} lies in the rank structure of the solution given our tensor decomposition in Equation (\ref{eq:tensor_decomp}).  If the rank of the problem explodes, \texttt{TTTT} compression and performance falters. In the worst case, a \texttt{TTTT} method will be less performant than a traditional $S_N$ method, due to the additional cost of working in the TT representation. On the other hand, if the rank of the problem is low, \texttt{TTTT} can provide large compressions and speedups.  One might then challenge \texttt{TTTT}, arguing that its applicability is only in regimes where we don't actually need large $N_\mathrm{angle}$; that is, if there is low rank structure, do we really need $S_N$ in the first place? Certainly, the 2D Gaussian diffusion problem demonstrating $\simeq10^4 \times$ speedups presented in \S\ref{sec:gauss2d} was purely academic---the diffusion approximation would have worked equally well, all the while promoting even better performance.  However, even in the 2D line source problem (\S \ref{sec:ls2d}), our ``worst-case scenario" for tensors, we still see large compressions and considerable speedups for the $N_\mathrm{angle} = 512 \times 1024$ calculation.  Moreover, at the final timestep, the ranks $[r_1,r_2] = [1460, 117]$ are significantly lower than ``full" rank $[r_{1,\mathrm{full}}, r_{2,\mathrm{full}}] = [500^2, 1024]$.  This indicates that at $\epsilon=10^{-4}$, \texttt{TTTT} can still find low rank structure in this challenging problem, even for a solution that is still not entirely angularly converged (recall slight deviations from analytics associated with $R_\mathrm{min}$).  Most interestingly, the crooked pipe test problem in \S \ref{sec:crooked} represents a scenario wherein there are both optically thin (inside pipe) and optically thick (outside pipe) regimes.  During the initial transient part of this problem, radiation $\sim$free streams through the optically thin pipe reaching its first obstacle, all the while heating the walls of pipe.  Getting this part of the problem correct is essential because its sets the stage for associated timescales for radiation propagation throughout the remainder of the pipe.  To accurately capture this phase, we need the most information from the angular structure of the radiation field, and \texttt{TTTT} naturally supplies it by a transient increase in rank, reaching a peak shortly into the evolution (accompanied by the lowest observed performance). Even still, at these times, the $N_\mathrm{angle} = 64 \times 128$ is $\sim 20\times$ faster.  All in all, diffusion-like methods are likely the right choice for the everywhere optically thick regime; perhaps ray tracing algorithms (see, e.g., references in \S \ref{sec:stellar}) are optimal for the very optically thin regime; but in the intermediate regime, or where both regimes reside within the same computational volume, $S_N$-like methods are often adopted, and this is where \texttt{TTTT} shines. 

\begin{table}[h]
    \centering
    \begin{tabular}{|c|c|c|c|c|}
        \hline
        Problem & $\epsilon$ & $N_\mathrm{angles}$ & Integrated Speedup & Minimum Compression \\ 
        \hline
        Gaussian Diffusion & $10^{-4}$ & $1024\times2048$ & $\simeq$37,000$\times$ & $\simeq$533,000$\times$ \\
        Line Source  & $10^{-4}$ & $2\times4$ & $\simeq$0.4$\times$ & $\simeq$2$\times$ \\
        Line Source  & $10^{-3}$ & $2\times4$ & $\simeq$0.4$\times$ & $\simeq$2$\times$ \\
        Line Source  & $10^{-4}$ & $32\times64$ & $\simeq$0.1$\times$ & $\simeq$3$\times$ \\
        Line Source  & $10^{-3}$ & $32\times64$ & $\simeq$0.3$\times$ & $\simeq$4$\times$ \\
        Line Source$^\star$  & $10^{-4}$ & $512\times1024$ & $\simeq$12$\times$ & $\simeq$290$\times$ \\
        Line Source$^\star$ & $10^{-3}$ & $512\times1024$ & $\simeq$41$\times$ & $\simeq$560$\times$ \\
        Crooked Pipe  & $10^{-3}$ & $2\times4$ & $\simeq$0.2$\times$ & $\simeq$2$\times$ \\
        Crooked Pipe  & $10^{-3}$ & $16\times32$ & $\simeq$3$\times$ & $\simeq$11$\times$ \\
        Crooked Pipe & $10^{-3}$ & $64\times128$ & $\simeq$34$\times$ & $\simeq$171$\times$ \\
        Stellar Irradiation$^\star$ & $10^{-4}$ & $512\times1024$ & $\simeq$62$\times$ & $\simeq$1054$\times$ \\
        \hline
    \end{tabular}
    \caption{Summary of the performance characterizations for the tests considered in this work. Columns present the problem, the rounding tolerance $\epsilon$, $N_\mathrm{angles}$, the integrated speedup (or slowdown when $<$1), and the minimum compression factor $\mathcal{C}$ achieved over the duration of the integration.  For the crooked pipe performance diagnostics, we only report measurements over the first 75,000 timesteps (i.e., $\simeq$25 ns simulation time) to match Figure \ref{fig:crooked_perf}. Problems with an asterisk superscript$^\star$ show projected speedups assuming no memory limitations; these problems cannot actually be run serially without tensor compression on our \texttt{Skylake} node testbed due to large $N_x N_y N_\theta N_\phi$.  All speedups are relative to our traditional $S_N$ \texttt{Python} reference metric (vacuum transport: $4.3\times10^7$, scattering: $3.2 \times 10^7$, absorption/emission: $3.1 \times 10^7$ angle updates per second).}
    \label{tab:perf_numbers}
\end{table}

Table \ref{tab:perf_numbers} provides an overview of the performance metrics taken from our presented test problems.  Rather than supplying instantaneous performance measures (as in Figures \ref{fig:gaussian2d}, \ref{fig:ls2d_perf}, and \ref{fig:stellar}), we now report the integrated speedups (i.e., a direct probe of the time to solution).  Despite the $\epsilon = 10^{-4}$, $N_\mathrm{angles} = 512 \times 1024$ line source problem presenting a ZCPS measure only 3$\times$ faster than traditional methods in its final timestep, we find that accelerations in earlier evolution reduce the overall time to solution by $\simeq$12$\times$ compared to traditional $S_N$.  Table \ref{tab:perf_numbers} also presents the minimum compression factor $\mathcal{C}$ achieved over the course of the integration.  In many cases, \texttt{TTTT} offers orders of magnitude compressions; we highlight problems (asterisk superscripts) wherein tensors enable otherwise impossible serial calculations on our \texttt{Skylake} testbed (due to memory requirements).  Albeit we still present speedup measures in these cases, they are only hypothetical and operate under the large assumption that traditional methods maintain the simple constant performance metric used elsewhere.

\texttt{TTTT} does not come without added complexity.  As shown in \S \ref{sec:hohlraum}, tensors introduce an \textit{additional axis} for global error convergence: the rounding tolerance $\epsilon$.  Converged solutions in both space and angle place a problem-dependent requirement on $\epsilon$.  Looking forward, as we move towards implicit algorithms invoking \texttt{TTTT}, iteration residual tolerances might interplay with $\epsilon$ in unexpected ways.  Most importantly, \texttt{TTTT} introduces a new trade space between errors.  In \S \ref{sec:ls2d}, we showed the line source solution at two different rounding tolerances, with the $\epsilon = 10^{-3}$ solution showing numerical ringing.  Whether the nature of this error is ``worse" than the error associated with ray effects at lower angular fidelity will likely be a problem dependent decision.  All the while, conservation of integrated radiation energy density is shown to be sensitive to the rounding tolerance, and we expect that the positivity of $I$ may be equally sensitive.  \texttt{TTTT}, as formulated, makes no guarantees as to the positivity of $I$ and enforcing positivity via a simple floor $\mathrm{max}(I, 0)$ is not straightforward in tensor format---alternative approaches might consider evolving $\sqrt{I}$, but this is outside the scope of this work.  

Finally, we note that this work has assumed a static background medium (i.e., $\mathrm{material \; velocity} = 0$) in Minkowski space.  In relativistic transport \citep[e.g.,][]{zhang21,white23}, it is possible that the evolved specific intensity (i.e., the lab-frame specific intensity or the ``tetrad"-frame specific intensity) may not exhibit low rank structure, even when the fluid-frame specific intensity would.  Instead evolving the fluid-frame specific intensity comes with its own numerical challenges, and we leave this concern for future works.  

\section{Conclusions} \label{sec:conclusions}
We have presented the \texttt{TTTT} algorithm for integrating the gray thermal transport equation via tensor networks.  We find that the specific intensity solution rank (1) determines compression and performance and (2) is highly problem dependent.  Nevertheless, in many cases, we find that \texttt{TTTT} gives large compressions and speedups relative to more traditional approaches to $S_N$, sometimes even enabling calculations that were otherwise impossible due to memory requirements.  Given these results, we find that tensor networks provide an exciting avenue for problems in thermal transport for regimes wherein $S_N$ is typically applied.

The tensor decomposition selected in this work was a choice motivated by (1) how the radiation source term typically invokes moments of the radiation field (right hand side of Equation \ref{eq:transport_tt}) and (2) how the gray thermal transport equation couples to the fluid via its moments (Equation \ref{eq:temp_eq}).  As shown in Equation (\ref{eq:tt_energy_density}), moments of the specific intensity are straightforwardly calculable given our decomposition.  However, other tensor decompositions exist.  Adopting a unique TT-core for each spatial axis might promote larger compressions and/or speedups, but also invites more severe explosion of rank for problems with complex spatial geometries.  Alternative tensor decompositions might benefit problems wherein angular complexity varies with position.  In our current decomposition, we expect that ``interactions" between space and angle invite rank growth.

In the context of block adaptive mesh refinement \citep[block AMR, see, e.g.,][]{mignone12,stone20,grete22,stone24}, block-local tensor decompositions might promote a path to discouraging interactions between space and angle if the angular structure of the radiation field is largely uniform throughout the block.  However, communication between different tensor solutions across different blocks is a challenging problem.  Overall, mapping \texttt{TTTT} to block AMR meshes demands significant attention.   

In this work, we have attempted to demonstrate the utility of tensors in thermal transport to a broad class of operators.  We demonstrated how to interact with various (1) flux stencils (including the complicated \cite{jiang21} asymptotic preserving flux) and (2) boundary conditions.  Yet still, we made several simplifications; e.g., the use of piecewise constant reconstruction and RK1 with simple operator splitting.  The introduction of high order reconstructions promotes non-linearities that require special handling in tensor arithmetic.  More complicated IMEX algorithms require careful consideration as to when rounding operations can/should occur throughout the integrator substages.  

Finally, shifting from gray thermal transport to multigroup provides yet another exciting future avenue for tensors.  Including frequency adds an additional dimension to the transport equation, exacerbating the curse of dimensionality for an $S_N$ approach and offering an additional avenue of compression for tensor trains. In vacuum, the transport operator is identical for each frequency group, therefore, we expect nearly perfect compression for problems not invoking the radiation source term.  Determining the rank structure for problems invoking frequency-dependent opacities will more generally determine the gains available through \texttt{TTTT} for multigroup thermal transport.  

\begin{acknowledgments}

A.A.G and A.D were supported by the Los Alamos National Laboratory (LANL) under the project ``Algorithm/Software/Hardware Co-design for High Energy Density applications'' at
the University of Michigan. A.D. was also supported, in part, by AFOSR Computational Mathematics Program under the Award \#FA9550-24-1-0246 to study the effects of randomized rounding in the present setting.
P.D.M, J.C.D, C.D.M., J.M.M., and L.F.R. were supported by LANL under the M$^3$AP project.  This research used resources provided by the Darwin testbed at LANL which is funded by the Computational Systems and Software Environments subprogram of LANL's Advanced Simulation and Computing program (NNSA/DOE). LANL is operated by Triad National Security, LLC, for the National Nuclear Security Administration of U.S. Department of Energy (Contract No. 89233218CNA000001). This document is approved for unlimited release under LA-UR-25-22716.
\end{acknowledgments}

\clearpage
\appendix

\section{Rounding} \label{sec:appendix_rounding}
Rounding is an essential component of a step-and-truncate tensor network solver to mitigate rank growth that would otherwise make the scheme not viable. In this section we consider a $d$ dimensional tensor with mode sizes $n$. 

\begin{figure}[htb]
    \centering
    \includegraphics[width=\textwidth]{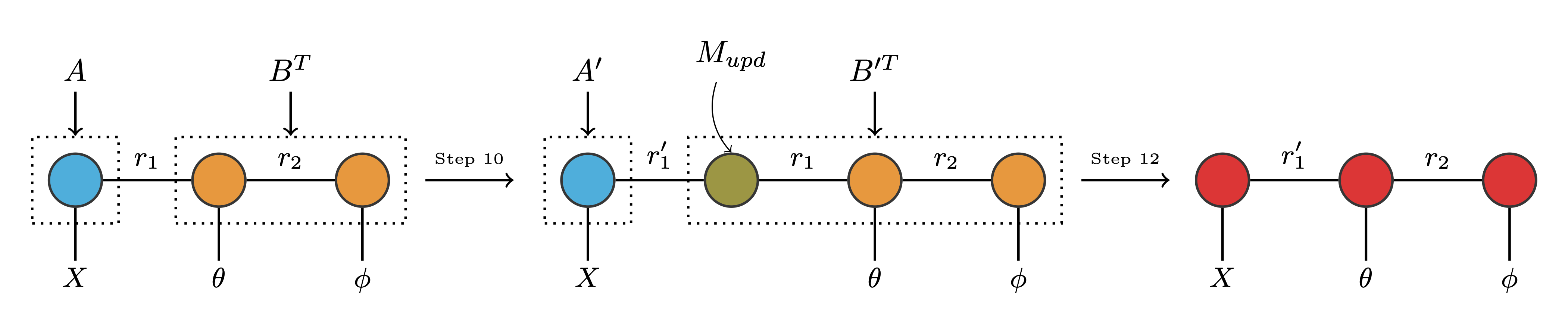}
    \caption{The figure illustrates the steps involved in a single iteration of the rounding algorithm. The first step involves defining $A$ and $B^T$ (left). The first TT rank is then truncated using one of the 3 algorithms discussed in the upcoming subsections. The update matrix $M_{upd}$ is absorbed into the next TT core through contraction along edge denoted by $r_1$ to get a TT with its first rank truncated from $r_1$ to $r'_1$.}
    \label{fig:rounding}
\end{figure}

Algorithm 1 describes a generalized TT-rounding procedure. The input to this procedure is the TT cores of an existing approximation, and a target set of ranks or accuracies. Figure \ref{fig:rounding} schematically describes a single iteration. The generalization takes on specific formulations for choices of \textbf{Step \ref{algline:10}} of the algorithm. In rounding algorithms, the \texttt{reshape} operator, similar to the functions provided by libraries like \texttt{NumPy}, is used to modify the shape of a tensor while ensuring that the total number of elements remains unchanged. 
Moreover, for any 2 tensors $\mathcal{X}$ and $\mathcal{Y}$, we use $\mathcal{X} \times_m^n \mathcal{Y}$ to denote the tensor contraction of $\mathcal{X}$ and $\mathcal{Y}$ along modes $m$ and $n$ respectively.

\begin{algorithm}
\SetAlgoLined
\KwIn{TT cores \( \{\mathcal{G}_i\}_{i=1}^d \) with ranks $\{r_k\}_{k=1}^{d-1}$ and target ranks $\{r'_k\}_{k=1}^{d-1}$ or target accuracy $\varepsilon$ (See Note on rank truncation below)}
\KwOut{Rounded TT-cores}

$\{$Backward sweep/Preprocessing phase$\}$

$B_{d-1}^T := \mathcal{G}_d$

\For{$k \gets d-2$ \KwTo $1$} 
{

    $B_k^T := \texttt{reshape}(\mathcal{G}_{k+1} \times_{-1}^1 B_{k+1}^T, (r_k, -1)) \label{algline:4}$
}

$\{$Forward sweep/Truncation phase$\}$

\For{$k \gets 1$ \KwTo $d-1$}{
    $A_k := \texttt{reshape}(\mathcal{G}_k, (r'_{k-1} N_k, r_k)) \label{algline:8}$

    $\{$Get rank $r'_k$ approximation of $A_k B_k^T$$\}$
    
    $[A_k', M_{upd}] = \texttt{truncate}(A_k, B_k, r'_k, \varepsilon) \label{algline:10}$
    
    $\mathcal{G}_k := \texttt{reshape}(A_k', (r'_{k-1}, N_k, r'_k)) \label{algline:11}$

    $\mathcal{G}_{k+1} :
    = M_{upd} \times_{-1}^1 \mathcal{G}_{k+1} \label{algline:12}$
}
Return $\mathcal{G}_1, \dots, \mathcal{G}_d$
\caption{TT-Rounding Meta-Algorithm}
\label{alg:ttround}
\end{algorithm}

We now describe several steps of this algorithm.
\begin{itemize}[label=$\bullet$, itemsep=1em]
    \item \textbf{Define $A_kB_k^T$} (Step \ref{algline:4}, \ref{algline:8}): An initial rank $r_k$ matrix product $A_kB_k^T$ is implicitly defined through $A_k$ and $B_k$. Here, $A_k \in \mathbb{R}^{m \times r_k}$, $B_k \in \mathbb{R}^{n \times r_k}$, $m = r'_{k-1}N_k$, and $n = \prod_{i=k+1}^{d} N_i.$
        
    \item \textbf{Rank truncation} (Step \ref{algline:10}): In this step $A_kB_k^T$ is approximated by rank $r'$ matrix product $A_k'B_k'^T$. $A_k' \in \mathbb{R}^{m \times r'_k}$, $B_k' \in \mathbb{R}^{n \times r'_k}$, $r'_k < r_k$, $m = r'_{k-1}N_k$, and $n = \prod_{i=k+1}^{d} N_i$. Additionally, $B'^T_k$ is obtained in the form of $M_{upd}B^T_k$. 
    
    \vspace{0.1cm}
    
    \textbf{\underline{Note}:} The truncation rank $r'_k$ could be stated explicitly by the user or could be implicitly determined based on user-specified error tolerance. If it is truncated by error, the error must be carefully balanced across dimensions so that overall relative error is guaranteed~\cite{oseledets2011tensor}. \label{note:rankerror}
    \footnote{The randomized rounding procedure is an exception to this. The randomized method discussed in this paper operates with a predetermined target rank rather than an error tolerance.} 
    This is the step that differentiates the 3 rounding algorithms discussed later. 

    \item \textbf{Form TT-core $\mathcal{G}_k$} (Step \ref{algline:11}): Matrix $A_k'$ acquired in the previous step is used to form the rounded TT-core $\mathcal{G}_k$ with reduced TT-ranks.

    \item \textbf{Absorbing $M_{upd}$ into the next core} (Step \ref{algline:12}): While the current core is formed using $A_k'$, we need to propagate the information in $B_k'^T$ into the next core using $M_{upd}$.
\end{itemize}

\noindent
In the following subsections, we will discuss some common approaches to tackle \textbf{Step \ref{algline:10}} of the above algorithm. Specifically, given $A, B$, and target rank $r'$, we will see how the \texttt{truncate} function for each approach returns a low-rank product in the form of $A' B'^T$, where $B'^T$ is expressed as an updated version of the original matrix $B^T$, represented in the form $M_{upd}B^T$.

\subsection{TT-SVD}
The traditional rounding approach introduced by \cite{oseledets2011tensor} is based on orthogonalizing a tensor train through successive QRs and compressing the cores using SVDs. Let $X = AB^T$. Also let $A = Q_A R_A$, and $B = Q_B R_B$
\footnote{While implementing, the QR decompositions for $B$ are performed during a preprocessing phase (or backward sweep).} 
represent the QR decompositions of $A$ and $B$ respectively. As a result, we have that

\begin{align}
    AB^T &= Q_A \underbrace{(R_A R_B^T)}_{R_{AB}} Q_B^T.
\end{align} 

We take a truncated SVD, to a given tolerance or rank bound, of $R_{AB}$ so that $R_{AB} \approx \tilde{U} \tilde{\Sigma} \tilde{V}^T$. Then we have 

\begin{align}
    X = AB^T &\approx (Q_A \tilde{U}) (\tilde{\Sigma} \tilde{V}^T Q_B^T) = A'B'^T,\\
    B'^T &= \tilde{\Sigma} \tilde{V}^T Q_B^T = (R^{-1} \tilde{V} \tilde{\Sigma})^T  B^T = M_{upd}B^T,\\
    A^\prime &= Q_A \tilde{U}.
\end{align}

\subsection{Gram-SVD}
The TT-SVD algorithm relies on an expensive orthogonalization step that involves a series of QR decompositions, which becomes the primary bottleneck not only in the rounding process but also in the overall solver. Due to the computational cost of this approach, there is a need for a more efficient alternative. Gram-SVD-based rounding, as proposed by \cite{al2022parallel} in the context of TTs, offers an elegant solution to this challenge. It involves working with Gram matrices of $A$ and $B$. It exploits the fact that the singular values of a matrix are the square roots of the eigenvalues of its respective Gram matrix. 

Let $G_A = A^TA$ and $G_B = B^TB$
\footnote{All Gram matrices of $B$ are evaluated during the preprocessing phase for efficient implementation.}
denote Gram matrices of $A$ and $B$ respectively. Additionally, let $G_A = V_A \Sigma_A^2 V_A^T$ and $G_B = V_B \Sigma_B^2 V_B^T$ represent the eigendecompositions of $G_A$ and $G_B$ respectively. Then we have 

\begin{align}
    X &= AB^T = A(V_A \Sigma_A^{-1} \Sigma_A V_A^T V_B \Sigma_B \Sigma_B^{-1} V_B^T) B^T, \\
     &= (A V_A \Sigma_A^{-1}) 
    \underbrace{(\Sigma_A V_A^T V_B \Sigma_B)}_{M} 
    (B V_B \Sigma_B^{-1})^T. \label{eq:gramfinal}
\end{align}

We take a truncated SVD of $M$; $M \approx \tilde{U} \tilde{\Sigma} \tilde{V}^T$. Substituting $M$ into \eqref{eq:gramfinal}, we arrive at

\begin{align}
    X &= AB^T \approx (A V_A \Sigma_A^{-1} \tilde{U}) (B V_B \Sigma_B^{-1} \tilde{V} \tilde{\Sigma})^T = A'B'^T,\\
    B'^T &=  (V_B \Sigma_B^{-1} \tilde{V} \tilde{\Sigma})^T B^T = M_{upd} B^T,\\
    A^{\prime} &= A V_A \Sigma_A^{-1} \tilde{U}.
\end{align}

\subsection{Randomized}
Randomized methods for rounding also offer significant potential for reducing computational costs. In the context of TTs, \cite{al2023randomized} explore randomized rounding, which utilizes randomized sketching to map a matrix onto a low-rank subspace. 
Consider the matrix $X \in \mathbb{R}^{m \times n}$. Projecting $X$ to a low-rank subspace translates to finding an orthogonal basis $Q \in \mathbb{R}^{m \times r}$ ($r < m, n$) such that $X \approx Q Q^T X$. The quality of this approximation is determined by the norm of the error given by $||(I - QQ^T) X||$. In the context of TTs, we can set $X = AB^T$. We need to find a $Q$ such that

\begin{align}
    (AB^T) &\approx Q Q^T (AB^T).\\
    X = AB^T &\approx (Q)(Q^TAB^T) = A'B'^T \label{eq:randfinal},\\
    B'^T &= (Q^T A)B^T = M_{upd} B^T,\\
    A' &= Q.
\end{align}
The steps above outline a general strategy to approximate a matrix ($X$) by projection to a low-rank subspace. Note that no randomized approach has been mentioned yet. The randomization comes into play during the evaluation of $Q$. More specifically, we use randomized sketching to find a low-rank subspace for $A$ and find its orthogonal basis using a QR factorization of the sketched subspace. We define,
\
\begin{align}
    W &= B^T \Omega \label{eq:randprojection},
\end{align}
 where $\Omega \in \mathbb{R}^{n \times r'}$ is a random Gaussian matrix and $r' < r$ is the target rank. Now, let $W = QR$ represent the QR decomposition of $W$. This $Q$ matrix is used in \eqref{eq:randfinal} to obtain the projection of $AB^T$ to a rank $r'$ subspace spanned by the columns of $Q$. Again, it is important to mention that the operation in \ref{eq:randprojection} is carried out during the preprocessing phase of the algorithm.

Note that in all of the low-rank matrix approximation methods discussed above, the left factor $A'$ has orthogonal columns. 

\subsection{Tradeoffs}

\paragraph{\textbf{Cost}}
The initial orthogonalization phase in traditional rounding is the most computation-intensive step. Gram-SVD based rounding works with smaller-sized Gram matrices of $A$ and $B$ and avoids the initial orthogonalization completely, thus reducing the overall cost. Although traditional rounding and Gram-SVD rounding have the same order of time complexity, the Gram-SVD-based implementation seems to be much faster. This is because the dominating operation in terms of cost comprises successive matrix multiplications, a highly optimized operation in linear algebra packages. In randomized rounding, the initial orthogonalization phase is avoided altogether, generating a massive potential to reduce computational costs. Assuming the target rank is the same for all 3 methods, randomized rounding achieves the highest speedup followed by Gram-SVD based and traditional approaches. Table \ref{tab:roundingcost} presents the computational cost of the TT rounding algorithms, excluding lower-order terms for clarity.

\begin{table}[h]
    \centering
    \begin{tabular}{|c|c|}
        \hline
        Algorithms & Cost (FLOPs) \\ 
        \hline
        Traditional & $6dNr^3 + 6dNr'r^2$ \\
        Gram-SVD based & $4dNr^3 + 4dNr'r^2$ \\
        Randomized & $4dNr'r^2 + 6dNr'^2r + 2dNr' - 2r'^3/3$ \\
        \hline
    \end{tabular}
    \caption{Traditional, Gram-SVD, and Randomized rounding algorithms and their associated costs (in FLOPs).}
    \label{tab:roundingcost}
\end{table}

\paragraph{\textbf{Stability}}
The biggest limitation of Gram-SVD-based rounding is tied to stability issues introduced as a result of working with Gram matrices. The effect of this is that the singular values obtained using the eigenvalues of gram matrices have a precision of $\sqrt{\epsilon}$ relative to the biggest singular value.   Traditional and randomized methods avoid the additional stability issues faced in Gram-SVD-based rounding since there are no computations involving gram matrices.

\subsection{Summary}
Traditional rounding is stable but very expensive. Gram-based rounding provides a significant speedup relative to traditional rounding but can introduce stability issues in the solver. For the test problems we have considered, these stability issues did not arise. The randomized algorithm attempts to provide a compromised solution. The main limitation of the current randomized rounding algorithms for TT is that, unlike traditional and Gram-SVD rounding, the user can only pre-specify the target rank $l$ and not an error tolerance $\epsilon$. This target may not always be known a priori in a solution that is updated at every timestep. Although we have considered mitigation strategies, we have opted to only present the results for Gram-SVD rounding in this work, as the role of the rounding tolerance $\epsilon$ is a central discussion point surrounding the viability of tensor networks for gray thermal transport. Randomized rounding requires a dedicated study surrounding the effects of introducing a rank growth bound. Nevertheless, we expect to find significant performance gains in this domain in future works.

%

%
%
\clearpage
\bibliography{refs}{}
\bibliographystyle{aasjournal}



\end{document}